%% file: 00_main.tex
\documentclass[10pt,twocolumn,letterpaper]{article}

\usepackage{cvpr}

\usepackage{graphicx}
\usepackage{amsmath}
\usepackage{nicefrac}
\usepackage{epsfig}
\usepackage{amssymb}
\usepackage{enumitem}
\usepackage[normalem]{ulem}
\usepackage[ruled]{algorithm2e}
\usepackage{multirow}
\usepackage{array}
\usepackage{ctable}
\usepackage{makecell}
\usepackage[capposition=bottom]{floatrow}
\usepackage{comment}
\DeclareMathAlphabet{\mathcal}{OMS}{cmsy}{m}{n}
\usepackage{textcomp}
\usepackage{caption}
\usepackage{booktabs}
\usepackage{mathtools}
\usepackage{textpos}
\usepackage{minitoc}
\usepackage{booktabs}
\usepackage{color}
\usepackage{wrapfig}
\usepackage[accsupp]{axessibility}
\usepackage[symbol]{footmisc}

\definecolor{citecolor}{RGB}{34,139,34}
\usepackage[pagebackref=true,breaklinks=true,colorlinks,
citecolor=citecolor,bookmarks=false]{hyperref}

\newcolumntype{P}[1]{>{\centering\arraybackslash}p{#1}}

\newcommand{\name}{\textsc{RealImpact}\xspace}
\newcommand{\objectfolder}{ObjectFolder\xspace}

\newcommand{\kleinpat}{\textsc{KleinPat}\xspace}
\newcommand{\neuralsound}{\textsc{NeuralSound}\xspace}

\newcommand{\myparagraph}[1]{\vspace{-4pt}\paragraph{#1}}
\newcommand{\precaption}{\vspace{-10pt}}
\newcommand{\pretablecaption}{\vspace{-5pt}}
\newcommand{\postcaption}{\vspace{-10pt}}
\usepackage[capitalize]{cleveref}
\crefname{section}{Sec.}{Secs.}
\Crefname{section}{Section}{Sections}
\Crefname{table}{Table}{Tables}
\crefname{table}{Tab.}{Tabs.}

\def\cvprPaperID{11902}
\def\confName{CVPR}
\def\confYear{2023}

\begin{document}

\title{\name: A Dataset of Impact Sound Fields for Real Objects 
}

\author{
Samuel Clarke\textsuperscript{1}\hspace{10mm} Ruohan Gao\textsuperscript{1}\hspace{11mm}  Mason Wang\textsuperscript{1} \hspace{9mm} Mark Rau\textsuperscript{1}\\ 
\hspace{4mm} Julia Xu\textsuperscript{1} \hspace{12mm} Jui-Hsien Wang\textsuperscript{2} \hspace{5mm} Doug L. James\textsuperscript{1} \hspace{8mm} Jiajun Wu\textsuperscript{1}\\
\textsuperscript{1}Stanford University \hspace{10mm} \textsuperscript{2}Adobe Research\\
}
\maketitle

\input{01_abstract}

\input{02_introduction}
\input{03_related_work}

\input{05_evaluation}
\input{04_dataset}
\input{06_benchmark}
\input{07_conclusion}

\paragraph{Acknowledgments.} This work is in part supported by Adobe, NSF CCRI \#2120095, NSF RI \#2211258, ONR MURI N00014-22-1-2740, AFOSR YIP FA9550-23-1-0127, the Stanford Institute for Human-Centered AI (HAI), Amazon, and Meta.

{\small
\bibliographystyle{ieee_fullname}
\bibliography{ref.bib}
}

\clearpage
\input{08_appendix}

\end{document}

%% file: 01_abstract.tex
\begin{abstract}
Objects make unique sounds under different perturbations, environment conditions, and poses relative to the listener. While prior works have modeled impact sounds and sound propagation in simulation, we lack a standard dataset of impact sound fields of real objects for audio-visual learning and calibration of the sim-to-real gap. We present \name, a large-scale dataset of real object impact sounds recorded under
controlled conditions. \name contains 150,000 recordings of impact sounds of 50 everyday objects with detailed annotations, including their impact locations, microphone locations, contact force profiles, material labels, and RGBD images.\footnote[1]{The project page and dataset are available at \url{https://samuelpclarke.com/realimpact/}} We make preliminary attempts to use our dataset as a reference to current simulation methods for estimating object impact sounds that match the real world. Moreover, we demonstrate the usefulness of our dataset as a testbed for acoustic and audio-visual learning via the evaluation of two benchmark tasks, including listener location classification and visual acoustic matching.

\end{abstract}

%% file: 02_introduction.tex
\section{Introduction}

Object sounds permeate our everyday natural environments as we both actively interact with them and passively perceive events in our environment.
The sound of a drinking glass bouncing on the floor assuages our fear that the glass would shatter. The click made by a knife making contact with a cutting board assures us that we have diced cleanly through a vegetable. And listening to the sound a painted mug makes when we tap it informs us of whether it is made of ceramic or metal.
What we perceive from sound complements what we perceive from vision by reinforcing, disambiguating, or augmenting it. 

Understanding the cause-and-effect relationships in these sounds at a fine-grained level can inform us about an object's material properties and geometry, as well as its contact and other environmental conditions.
Capturing these relationships from real-world data can help us improve our models toward more realistic physical simulations, with applications in virtual reality, animation, and training learning-based frameworks in simulation.

The sounds we perceive from objects are the result of many intricate physical processes: they encode important properties about the object itself (e.g., geometry, material, mechanical properties), as well as the surrounding environment (e.g., room size, other passive objects present, materials of furniture in the room). More specifically, when a hard object is struck, it vibrates according to its mass and stiffness, and the shape of the object determines the \emph{mode shapes} of the dominant vibration patterns (\S\ref{sec:modal_analysis}). Acoustic waves are then emitted into the medium, typically air, bouncing around in the room and interacting with surrounding objects and the room itself before reaching our ear or a microphone to be perceived as pressure fluctuations (\S\ref{sec:acoustic_transfer}).

Prior work has explored using physical simulation~\cite{james2006precomputed,wang2019kleinpat} or learning-based methods~\cite{jin2020deep,jin2022neuralsound} to reconstruct the sound generation process virtually, as well as building 3D environments with simulated spatial audio for embodied audio-visual learning~\cite{chen20soundspaces,gan2020threedworld,gao2020visualechoes,sagnik-eccv2022,senthil-iccv2021}. However, there has been little work on building physical apparatuses and feasible measurement process to quantify sounds made by the everyday objects, despite their importance and intimate relationship with our daily lives. As a result, the evaluations of the methods above are largely established on subjective metrics such as user studies. 

To address this gap, we introduce \name, a dataset containing 150k recordings of 50 everyday objects, each being struck from 5 distinct impact positions. For each impact point, we capture sounds at 600 field points to provide comprehensive coverage of the frequency components of the sounds and how they are distributed spatially. 

\name thus provides all the inputs most current simulation frameworks needed to simulate each sound, while also providing the ground truth recording for comparison. We show that \name can be used for various downstream auditory and audio-visual learning tasks, such as listener location classification (\S\ref{sec:listener_location_classification}) and visual acoustic matching (\S\ref{sec:visual_acoustic_matching}). These results demonstrate that sound fields can help improve machine perception and understanding of the world, and motivate further studies of even more accurate simulation methodologies to reduce the sim-to-real gap for future applications.

We make three contributions. First, we design an automated setup for collecting high-fidelity, annotated recordings of sounds by controlled striking of everyday objects. Second, using this setup, we acquire a large dataset of spatialized object sounds, \name. Third, we motivate the utility of \name by (a) using it to perform comparisons to results generated by current state-of-the-art sound simulation frameworks and (b) evaluating two benchmark tasks for acoustic and audio-visual learning.

%% file: 03_related_work.tex
\section{Related Work}
\begin{figure*}[t]
    \centering
    \includegraphics[width=\linewidth]{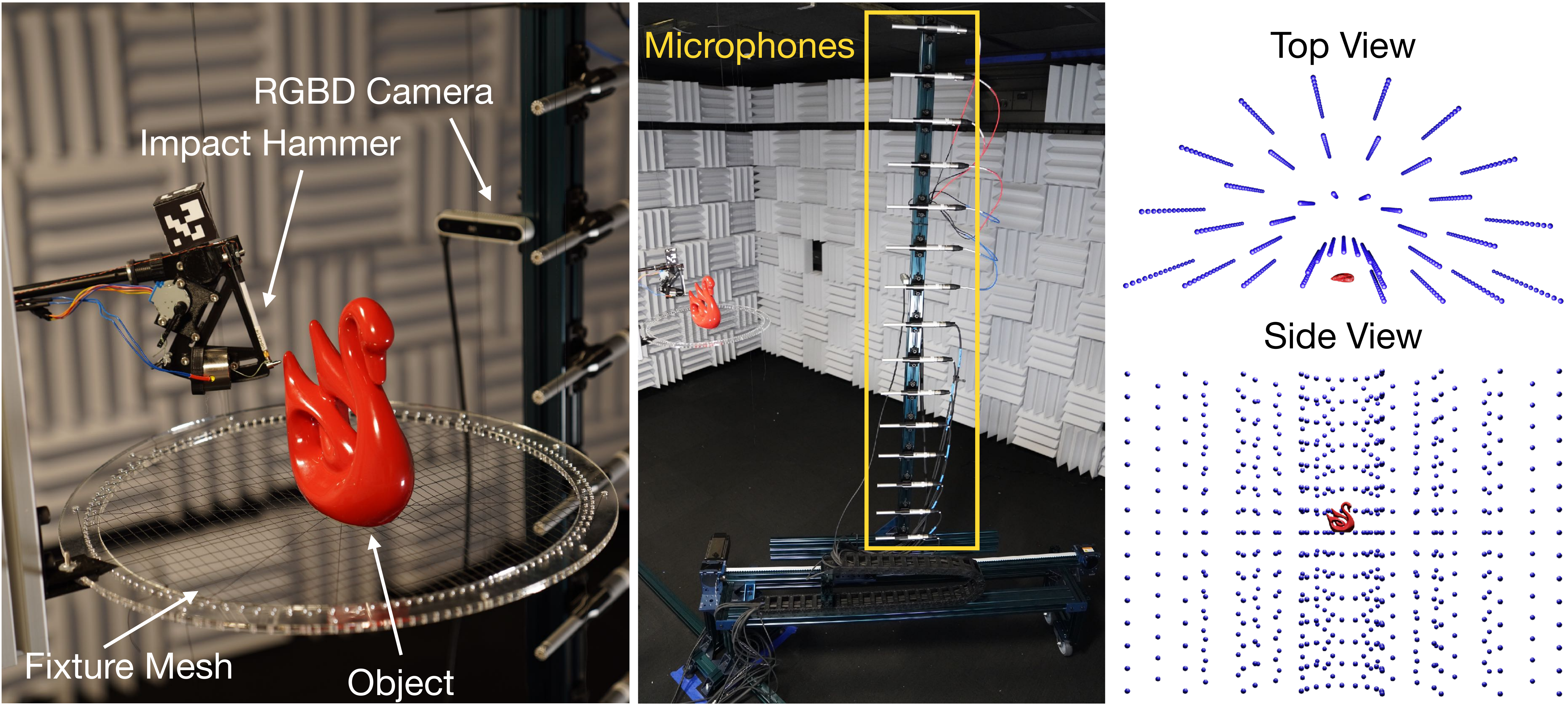}
    \caption{Pipeline for the acquisition of spatialized impact sounds: \textbf{(Left)} The object is placed at the center of the measurement platform and aligned with mesh threads.
    The impact hammer is positioned to strike a target vertex on the object. \textbf{(Center)} The gantry moves the microphones to 40 different positions within a semi-cylinder of the object, with the automated hammer mechanism striking the object to record the sound at each position. \textbf{(Right)} By the end of the recording process, for each of the 5 vertices of each object, recordings from 600 different microphone positions have been collected within the semi-cylinder to one side of the object, as shown.\postcaption}
    \label{fig:recording_illustration}
\end{figure*}

\myparagraph{Datasets of Object Sounds.}
Many datasets of object sounds have been introduced, each varying in the details of their collection, based on the applications they target. The \textit{Greatest Hits Dataset}~\cite{owens2016visually} includes audio-video recordings of thousands of impacts between real objects from the wild. The recordings were not taken in a controlled environment, and each impact is induced by a human with a drumstick, polluting each object's impact sound with the sound of the rather resonant drumstick. The \textit{Sound-20K Dataset}~\cite{zhang2017generative} is a fully synthetic dataset of 20,000 simulated recordings of objects being dropped in virtual environments. More recently, ObjectFolder~\cite{gao2021objectfolder,gao2022objectfolder} is introduced as a large dataset of trained implicit models for generating the sounds objects make when impacted at arbitrary locations. However, these are once again trained only on data from simulation, and they do not model the acoustic transfer properties of the objects, only their structural vibratory response.

\vspace{-0.1in}
\myparagraph{Physics-based Sound Rendering.}
Realistic sound rendering has been a long-held goal in computer music, interactive virtual environments, and computer animation~\cite{James:2016:PBS:2897826.2927375,Wang:2018:TWS:3197517.3201318,liu2020sound,smith2010physical}. By modeling the underlying physical processes of vibrations, the computer graphics community demonstrated convincing synthesized sounds for vibrating solids~\cite{OBrien:2001:SSF,obrien2002,james2006precomputed,Langlois:2014:EMC}, shells~\cite{chadwick2009}, rods~\cite{schweickart2017animating}, and even fluids~\cite{Langlois:2016:Bubbles,Wang:2018:TWS:3197517.3201318}. \cite{james2006precomputed,Wang:2018:TWS:3197517.3201318} further showed that it is important for high-quality sound rendering to capture the amplitude and spatial structure of radiating sound fields. However, computing these acoustic transfer fields is time-consuming as they are typically solved in the frequency domain, one frequency at a time. In \kleinpat~\cite{wang2019kleinpat}, the authors showed that by conflating multiple vibrating modes into one time-domain solve, getting all-frequency transfer maps can be done much faster, usually on the order of minutes. These models require careful simulation and material parameters tuning for the best results. To alleviate this issue, several works proposed to sample audio clips~\cite{ren2013example} and impulse responses~\cite{traer2019perceptually} to reconstruct the material definitions. Recently, a few works~\cite{jin2020deep,jin2022neuralsound} have explored using learning to approximate both the vibration and transfer computations using simulated training data. We provide timely real data that such simulations could use to validate their outputs and tune their performances.

\vspace{-0.1in}
\myparagraph{Recording Sounds Made by Real Objects.}
Whereas many learning-based frameworks have traditionally used simulation results as their ``ground truth'' for learning acoustic models of object vibrations and their transfer, some works have proposed to fit acoustic object models directly from data using digital signal processing with more relaxed model assumptions about rigid body vibrations.

Pai~\etal~\cite{pai2001scanning} describe a framework for scanning physical objects across multiple modalities, measuring visual, tactile, and audio properties of some everyday objects. They fit a data-driven acoustic model based on modal vibration for an object by striking it at different points and recording the ensuing sound from a single position per impact point, assuming constant acoustic transfer across the object. DiffImpact~\cite{clarke2021diffimpact} similarly fits modal models to real recordings of objects, but assumes a constant modal response and transfer across the object since their data lacked annotations of the impact point and microphone location. Perhaps most similar to our work, Corbett~\etal~\cite{corbett2007timbrefields} collect recordings of striking an object at three different impact points, positioning a microphone at 19 different positions per impact point. We collect recordings from an order of magnitude more microphone positions to empirically demonstrate that acoustic transfer varies rather drastically over a much finer resolution than can be captured at 19 different locations. 

Also, while these prior works have collected datasets from real audio, none have publicly released their datasets. Furthermore, since many simulation frameworks are designed to simulate audio of objects vibrating freely in an anechoic space, the recordings collected by these works are unsuitable for a fair comparison, since they are recorded with objects in contact conditions like resting on tables or grasped in hands, which greatly hinder free vibrations. In contrast, we propose a novel capture system where objects rest on a thread mesh in an acoustically treated room, which more closely approximates free vibration in an anechoic environment.

Finally, from outside the domain of object impact sounds, Bellows~\etal~\cite{bellows2018spherical} have an extensive project measuring the sound directivity of musical instruments while being played by musicians. The measurements take place in a large anechoic chamber and are recorded with a rotating semi-circular microphone array resulting in 2,522 unique microphone positions. The raw measurements are not provided, but the directivity patterns are available as spherical harmonic decompositions~\cite{bellows2022directivity}.

\vspace{-0.1in}
\myparagraph{Visual Learning of Sounds in Space.} 
Both audio and visual data convey crucial spatial information. 
Recent inspiring works have explored many interesting tasks connecting visual learning and sound in 3D space, including visually-guided audio spatialization~\cite{gao2019visualsound,morgadoNIPS18,garg2021geometry,zhou2020sep,xu2021visually}, sound source localization in video frames~\cite{chen2022visual,arandjelovic2017look,Senocak_2018_CVPR}, learning audio-visual spatial correspondence~\cite{chen2022sound,yang2020telling}, and building audio-visual 3D environments~\cite{chen20soundspaces,gan2020threedworld} for an array of embodied learning tasks~\cite{chen2021waypoints,gan2020look,dean,gao2020visualechoes,sagnik-eccv2022,senthil-iccv2021}. We show how our dataset can be used to evaluate real-world performance of auditory and audiovisual
learning frameworks on two novel tasks.

%% file: 05_evaluation.tex
\section{Physics-Based Sound Synthesis}

We begin with some background about physics-based sound simulation for rigid objects to motivate design choices for our dataset and provide context for our baseline simulation frameworks and their parameters. Here we briefly summarize a commonly used sound synthesis pipeline for rigid objects. For a more detailed introduction, we recommend the article from James~\etal\cite{James:2016:PBS:2897826.2927375}.

\subsection{Modal Sound Synthesis}
\label{sec:modal_analysis}

\def\u{\boldsymbol{u}}
\def\q{\boldsymbol{q}}
\def\f{\boldsymbol{f}}

When a contact force is applied to an object (e.g., your dinner plate hits the dishwasher handle), depending on the location of contact, various vibration modes can get excited and eventually die down due to internal damping. Mathematically, the vibration's displacement vector $\u(t)$ can be low-rank approximated as
\begin{equation}
\u(t) = U \q(t) = [\hat{\u}_1 \cdots \hat{\u}_K] \q(t),
\end{equation}
where $U$ is the modal matrix with mode shapes $\hat{\u}_i$ and $\q(t) \in \mathbb{R}^{K}$ the modal coordinates. The equations of motion are
\begin{equation}
M \ddot{\u} + C \dot{\u} + K \u = \f, \label{eq:equation_of_motion}
\end{equation}
where $M$, $C$, and $K$ are the mass, damping, and stiffness matrix, respectively\footnote{Formulas of how to compute these matrices can be found in~\cite{obrien2002}.}, and $\f$ is the external force vector. It is typically assumed in the literature that the damping can be approximated by Rayleigh damping, $C = \alpha M + \beta K$; with this convenient assumption, Eq.~\eqref{eq:equation_of_motion} can be re-written in the subspace defined by $U$ as
\begin{equation}
\ddot{\q} + (\alpha \mathbf{I} + \beta \Lambda) \dot{\q} + \Lambda \u = U^\intercal \f, \label{eq:equation_of_motion_subspace}
\end{equation}
where $\mathbf{I}$ is the identity matrix and $\Lambda = \text{diag}(\omega_1^2, ..., \omega_K^2)$ is a diagonal matrix of involving angular frequencies $\omega_i$. Since the damping can significantly affect material perception~\cite{klatzky2000perception}, the Rayleigh damping can potentially model real-world objects poorly. In addition, there are two scalar properties $\alpha$ and $\beta$ to fit, and in previous work, these are typically hand-picked to produce sounds that are closest to a given material. Also note that this formulation is based on linear modal analysis~\cite{shabana2012theory,shabana2013dynamics}, which assumes the vibrations are infinitesimal, or, in other words, the object is approximately rigid.

\subsection{Acoustic Transfer}
\label{sec:acoustic_transfer}

Sound radiates from an object's surface into the surrounding medium as pressure waves. Since the modes decay slowly over time, it is convenient to work in the frequency domain~\cite{james2006precomputed}. The distribution of the wave magnitudes in space, $\hat{p}(\boldsymbol{x}; \omega)$, is referred to in the literature~\cite{james2006precomputed} as the \emph{acoustic transfer} function. With a given set of vibrational boundary conditions, such as those given by Eq.~\eqref{eq:equation_of_motion_subspace}, one can solve the frequency-domain Helmholtz equation~\cite{james2006precomputed,chadwick2009,Langlois:2014:EMC} using boundary element methods~\cite{ciscowski1991,gumerov2005,Bebendorf2000} or a time-domain wave equation before reconstructing the acoustic transfer fields~\cite{wang2019kleinpat}. To display the acoustic transfer fields at runtime, we may compress the representation using multi-point dipole sources~\cite{james2006precomputed} or single-point multipole sources~\cite{zheng10}. Another increasingly popular way of storing and displaying the field is to leverage the radial structure (or lack thereof) in the far-field radiation to store Far-field Acoustic Transfer (FFAT) maps, which are rectangular, image-like textures that capture the angular radiation pattern of a given object. FFAT maps can be quickly reparametrized (e.g., equalized, time-delayed) at runtime to display object impact sounds with user interactions~\cite{wang2019kleinpat}.

%% file: 04_dataset.tex
\section{The \name Dataset}

We introduce \name, a dataset of 150,000 real object impact sounds. Along with these sounds, we record the force profiles from the impact hammer we used to strike the objects, as well as an RGBD image of the object from each azimuth angle and radial distance from which the audio recording is measured.
Below, we introduce the hardware setup for collecting the data, the objects we use, and our data collection pipeline.

\subsection{Hardware Setup}
We collect all recordings in an acoustically treated room (see Appendix~\ref{app:environment} for additional details).
We designed a cylindrical gantry system for moving the microphones to precise positions in space, shown in Figure~\ref{fig:recording_illustration}. The gantry system moves a 1.82-meter-tall vertical column of 15 Dayton Audio EMM6 calibrated measurement microphones which are evenly and precisely spaced along the column. It moves this column precisely in two degrees of freedom: azimuth and distance, with a precision of 1$^{\circ}$ and 1~mm, respectively.
We suspended a mesh of polyester threads precisely at the axis of rotation of this gantry, centering it vertically along the column of 15 microphones. This mesh holds the objects in place while minimizing contact damping and maximizing the acoustic transparency of the surface holding the object. Furthermore, the layout of the mesh provides visual guidance for precisely positioning the objects in a repeatable manner.

To measure the acoustic transfer from the object to the microphones, the impact force needs to be recorded, allowing an input-output relation to be found. We used a PCB 086E80 impact hammer to strike each object. The impact hammer is incorporated into a custom automated striking mechanism, which strikes objects precisely and repeatedly while being as silent as possible. The mechanism uses a motor to wind the hammer back to contact an electromagnet; then, upon recording, the electromagnet releases the hammer. Actuating the electromagnet is completely silent, so the noise created by this mechanism is minimal during each strike. This mechanism is mounted on a microphone stand to be able to position it rigidly to strike objects at arbitrary locations. See Appendix~\ref{app:apparatus}~and~\ref{app:hammer} for additional details on the recording apparatus and hammer impacts, respectively.

The impact hammer has a calibrated force transducer in its tip, measuring contact forces at the same temporal resolution as our audio. The impact hammer and microphones are all read in a time-synchronized fashion by using two Motu 8M audio interfaces connected by an optical cable. Each audio interface has digitally-controlled amplifier gains, which must be tuned up or down for object sounds that are relatively quiet or loud, respectively, to boost the signal as much as possible while also preventing clipping. Because these gains are digitally controlled, we can record and adjust them in a precise and repeatable manner throughout our experiments. The recordings were made at a sample rate of 48000~Hz.

We also attach a RealSense D415 RGBD camera to the column, aligned with the first microphone above mesh-level, to take RGBD images with our audio measurements.

\subsection{Objects}

\begin{figure}[t]
    \center
    \includegraphics[width=\linewidth]{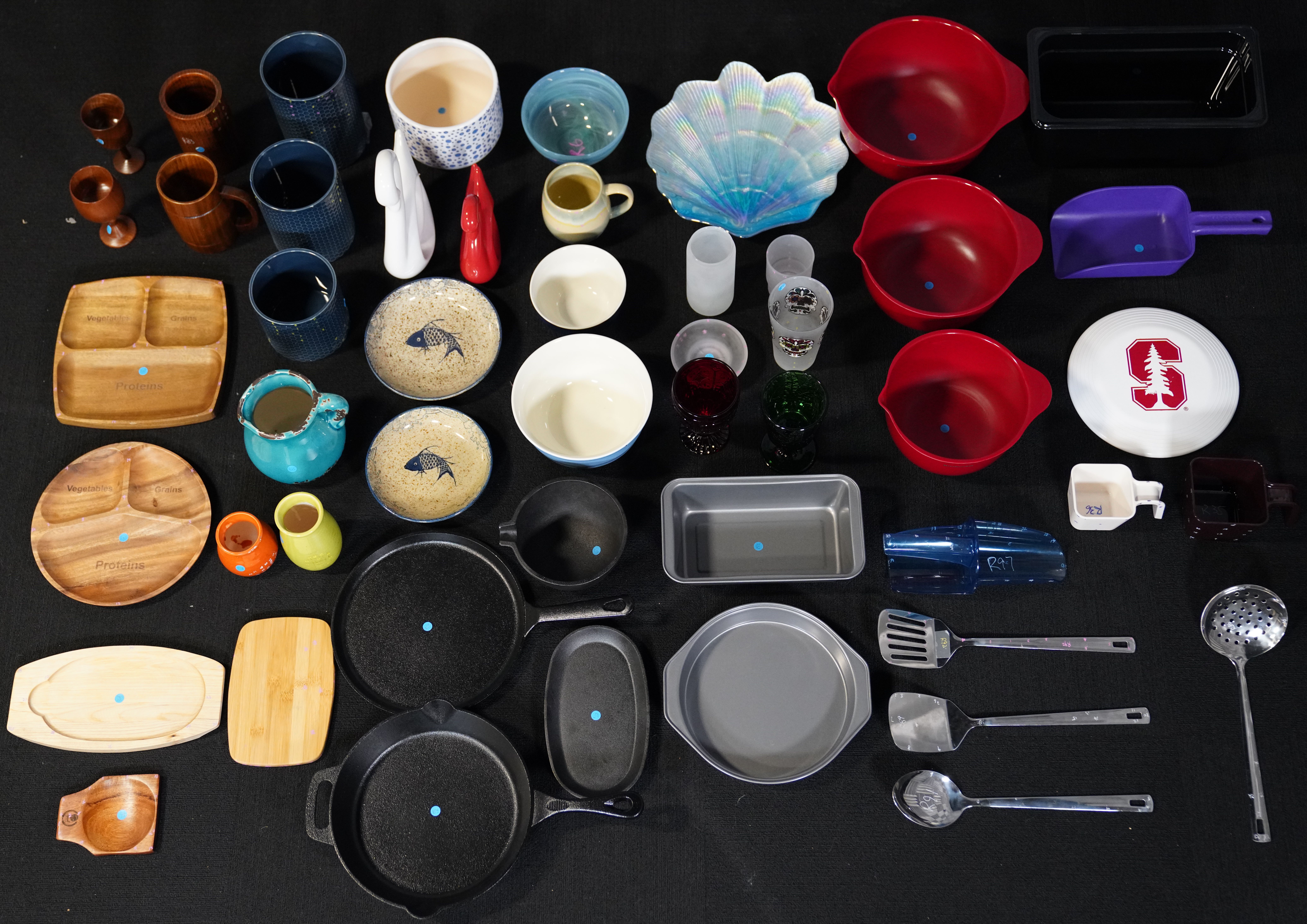}
    \caption{The real objects used in our dataset. Objects are clustered by material: (from top left) wood, ceramic, glass, plastic, (from bottom center) iron, steel, and polycarbonate.\postcaption}
    \label{fig:objects}
\end{figure}
We purchase 50 objects from the ObjectFolder dataset~\cite{gao2023objectfolder},
which is comprised of commonly used household objects like a ceramic mug, drinking cup, plastic bin, and wood vase. Each object in \name has a high-resolution 3D mesh model generated from a scan of the real object~\cite{gao2022objectfolder}, which can be used in simulation frameworks. We select objects which are rigid and consist of a single homogeneous material belonging to one of the following categories: ceramic, glass, wood, plastic, iron, polycarbonate, and steel---materials that have $\alpha$ and $\beta$ parameters available and widely used in the physics-based sound rendering literature~\cite{wang2019kleinpat,jin2020deep,jin2022neuralsound}. Figure~\ref{fig:objects} shows all objects used for data collection. These objects all have a scale and mass suitable for data collection using our hardware setup.

\subsection{Data Collection Pipeline}
\label{sec:data_collection}

\begin{figure}[t]
    \center
    \includegraphics[width=0.92\linewidth]{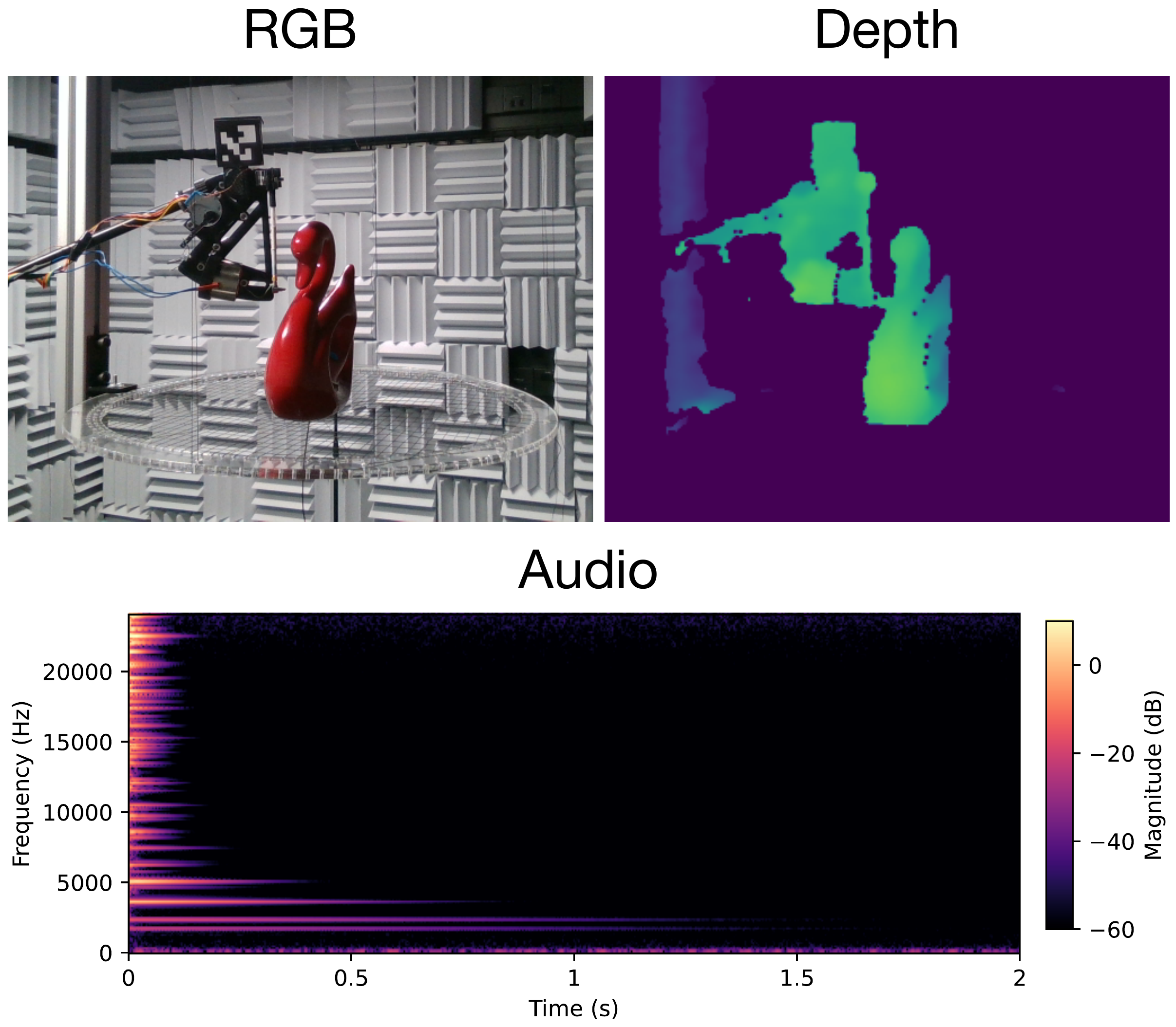}\precaption
    \caption{An RGBD image and audio recordings from all 15 microphones are collected at each gantry position for each vertex we impact on an object.\postcaption}
    \label{fig:modalities}
\end{figure}

For each object, we first place the object on the supporting mesh, matching the features of the mesh to the distinctive geometric features of the object to position it in a repeatable manner. We then select 5 vertices from the virtual mesh at which to strike the object. For each vertex, we first position the hammer mechanism to strike the vertex. Since our gantry collects recordings on a semi-cylinder to one side of the object, we position the hammer mechanism to the opposite side of the semi-cylinder both to not impinge the motion of the gantry and to minimize blocking acoustic radiation from the surface of the object toward the microphones. For each vertex, we move the gantry to 40 positions: a grid of 10 angles in 20-degree increments from 0 through 180, at 4 distances of 0.23, 0.56, 0.90, and 1.23 meters from the center of the mesh. We take an RGBD image at each position in addition to the audio recordings. A diagram of the microphone positions relative to an example object is shown in Figure~\ref{fig:recording_illustration}. An example of each of the modalities captured from one position is shown in Figure~\ref{fig:modalities}. See the Supplementary Materials for an example video of how an object is recorded.

\subsection{Processing}
\label{sec:processing}

The impact hammer strikes are not necessarily constant across measurements, but the discrepancy can be corrected since the force is measured. This is achieved by deconvolving the force signal from the microphone signal with frequency domain division as $m_c = \mathcal{F}^{-1}\left(\mathcal{F}\left(m \right)/\mathcal{F} \left(i \right)\right)$, where $i$ is the impact hammer signal, $m$ is the microphone signal, and $m_c$ is the corrected microphone signal. $\mathcal{F}$ and $\mathcal{F}^{-1}$ are the forward and inverse discrete Fourier transforms, respectively. The hammer signal is windowed such that only the samples within 1\% of the force peak are kept, and all other samples are deemed noise and set as zero, reducing noise in the corrected microphone signal.

To create transfer maps of the recordings, mode fitting is performed on each corrected microphone signal. The modes are fit using the method of~\cite{chowdhuryDAFx2019}. First, the vibrational frequencies are fit with a simple peak-picking algorithm performed on $\mathcal{F}\left( m_c \right)$. Decay rates are fit by bandpassing $m_c$ at the mode frequencies, applying a Root-Mean-Squared level detector, and using linear regression to estimate the slope of the energy envelope. The amplitudes are set as the magnitude of the mode frequency peak in $\mathcal{F}\left( m_c \right)$. Transfer maps are then formed for each vibrational frequency by displaying the magnitude at each measurement location with respect to rotation and height, as shown in Figures~\ref{fig:transfer_resolution} and~\ref{fig:transfer_repeatability}. 

\subsection{Validation}
\begin{figure}[t]
    \center
    \includegraphics[width=\linewidth]{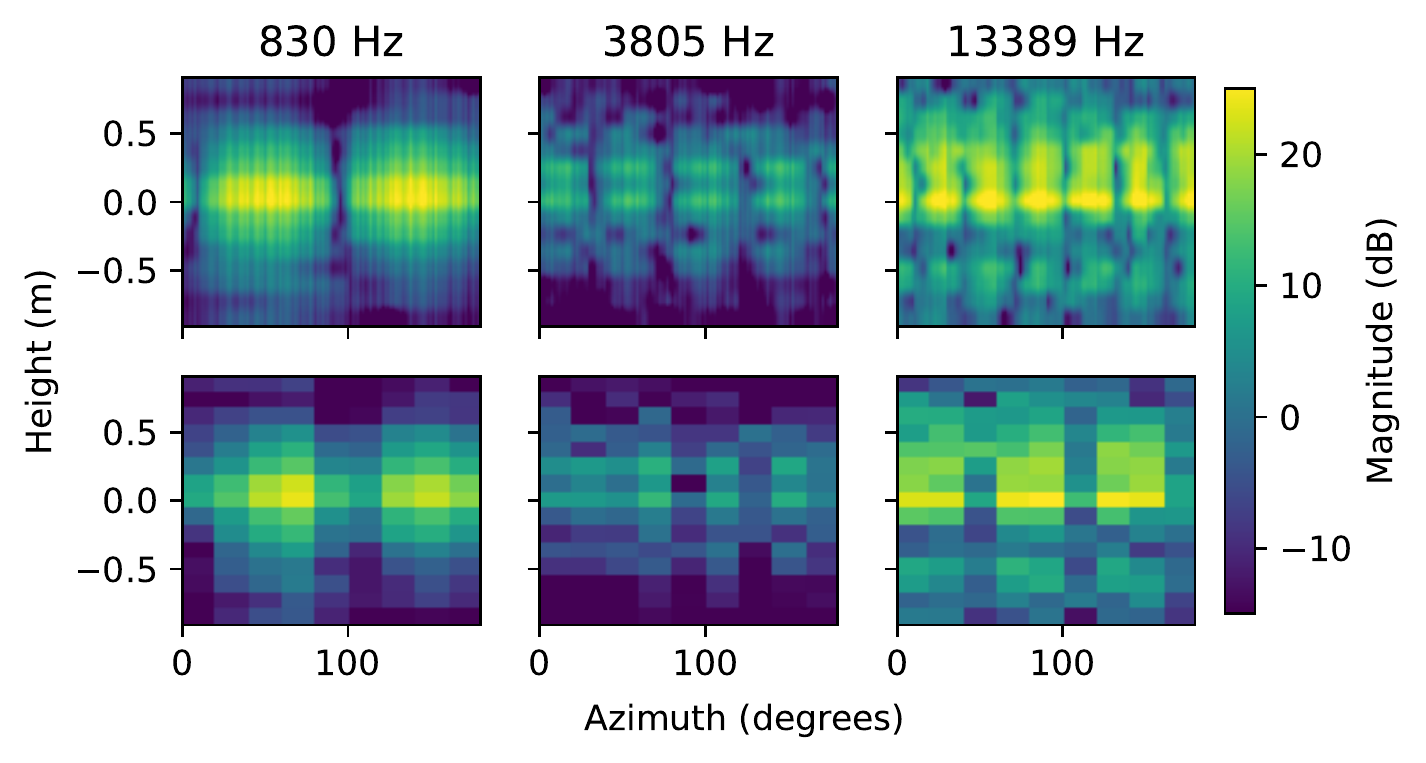}\precaption
    \caption{Comparing different azimuth resolutions for mode shape transfer maps of a ceramic bowl, measured at 23~cm from the center. The top row shows maps measured at a resolution of 1 degree, while the bottom row shows those measured at a resolution of 20 degrees.\postcaption}
    \label{fig:transfer_resolution}
\end{figure}

\myparagraph{Spatial Sampling.}
\label{sec:spatial_sampling}
We use a 20$^{\circ}$ resolution of azimuth angle for the spatial sampling as a compromise to reduce measurement time while still adding benefit for certain sound-related tasks. We take one set of measurements with 1$^{\circ}$ rotations on one of our objects (a ceramic bowl) as a comparison. Figure~\ref{fig:transfer_resolution} shows measured acoustic transfer maps for sample vibrational frequencies with  both 1$^{\circ}$ and 20$^{\circ}$ microphone rotations. The lowest-frequency mode shape varies gradually with the azimuth angle. But note that at the highest-frequency mode shape shown, the frequency of the repeating spatial pattern is beyond the Nyquist frequency of our azimuth sampling resolution. We show the implications of attempting to na\"{i}vely interpolate from these low-resolution transfer maps in Appendix~\ref{app:interpolation}.

\begin{figure}[t]
    \center
    \includegraphics[width=\linewidth]{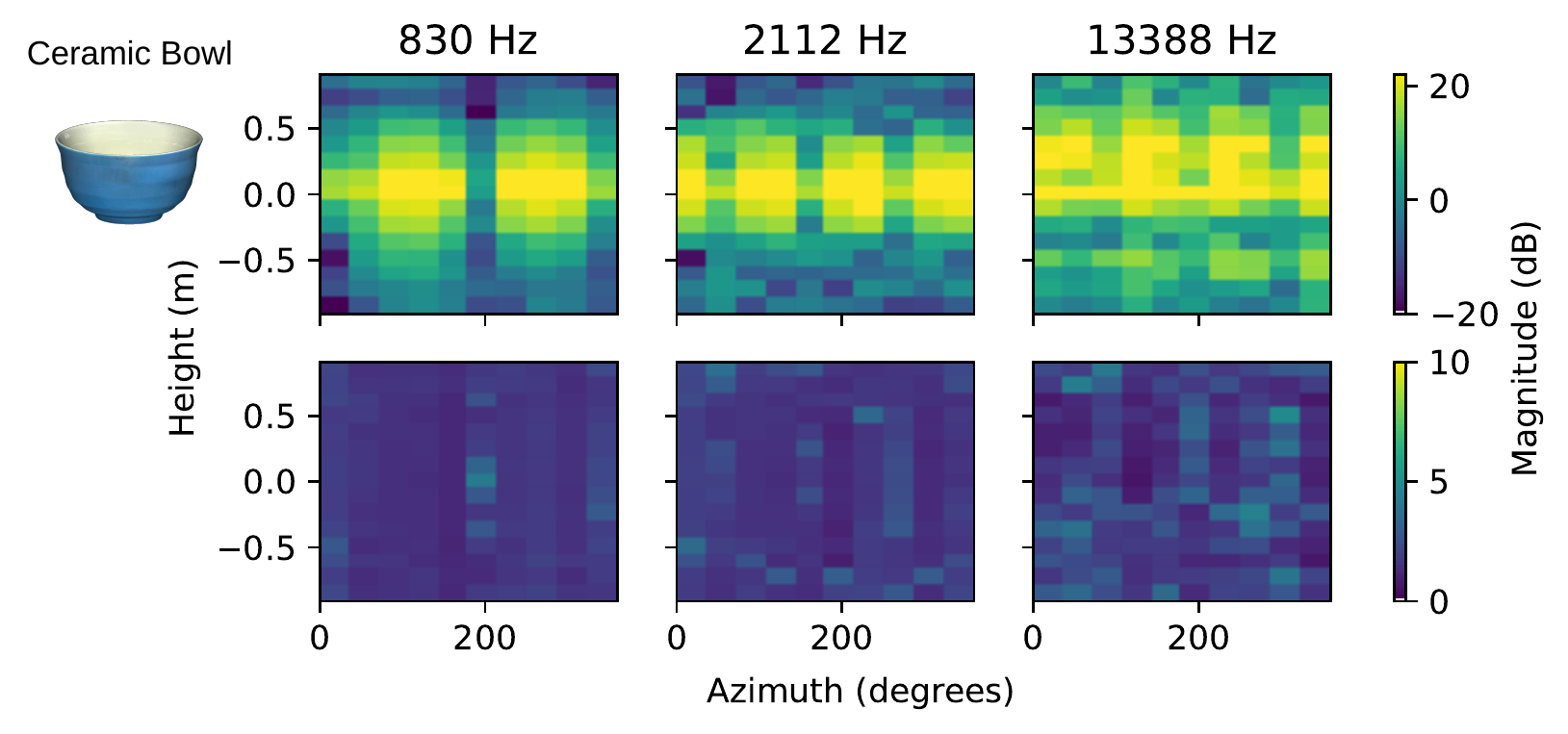}\precaption
    \caption{Measuring repeatability of our measurements by visualizing transfer maps of vibrational frequencies of the ceramic bowl, measured at 23~cm from the center. The top row shows the mean of 10 trials of measurements, while the bottom shows the relative standard deviation of the 10 trials.\postcaption}
    \label{fig:transfer_repeatability}
\end{figure}

\myparagraph{Repeatability.} 
\label{sec:repeatability}
We verify two aspects of the repeatability of our design: the repeatability of the gantry's position and the repeatability of our resulting audio measurements. While our gantry is capable of achieving high angular precision while it is controlled, we completely power it off during each recording, in order to eliminate motor and power supply noises from our recordings. During these periods, the wheels may settle into the carpet at a slightly different angle than we have commanded. We perform four trials of moving to the commanded angles and find that the maximum angular error across all trials did not exceed 1$^{\circ}$, with the mean error being 0.26$^{\circ}$. This translates to a maximum error of 2 cm in Cartesian space, only reached when the gantry is at its farthest distance from the center.

For the repeatability of our measurements, we conduct 10 trials of our measurements on the same ceramic bowl, striking the same target vertex and using the sample positions we used throughout our dataset. We show the mean and standard deviations of the transfers we measured at some sample vibrational frequencies in Figure~\ref{fig:transfer_repeatability}, with results of objects of additional materials in Appendix~\ref{app:repeatability}. Our results suggest that variations may be highest at the boundaries of nodes in the transfer map. At these locations, minor errors in the azimuth angle could cause significant changes in transfer measurement. Furthermore, at these positions, the signal is lower at this frequency, so the effects of noise can be more pronounced.

%% file: 06_benchmark.tex
\section{Applications}

In this section, we demonstrate some use cases of~\name with practical, multimodal applications.

\begin{table*}[t!]
\setlength{\tabcolsep}{5pt}
\begin{tabular}{lcccccc}
\toprule & \multicolumn{3}{c}{\name Deconvolved} & \multicolumn{3}{c}{\name Deconvolved + Denoised}\\
\cmidrule(lr){2-4}\cmidrule(lr){5-7}
 & L1 Spectral & Envelope ($\times 10^{-3}$)& CDPAM  & L1 Spectral & Envelope ($\times 10^{-3}$)& CDPAM  \\
\midrule
\textsc{White Noise}  & 4.68 & 9.54 & 1.38 & 5.22 & 9.87 & 1.39 \\
\textsc{Random Impact Sound}  & 0.728 & \textbf{4.17} & 0.121 & 0.150 & 4.97 & 0.0880 \\ 
\kleinpat~\cite{wang2019kleinpat}   &  \textbf{0.632}  & 4.63 & 0.117 & \textbf{0.0982} & \textbf{4.63} & 0.0975\\ 
\textsc{NeuralSound}~\cite{jin2022neuralsound}   &  0.673  & 23.0 & \textbf{0.102} & 0.133 & 22.8 & \textbf{0.0750} \\

\textsc{ObjectFolder 2.0}~\cite{gao2022objectfolder}    &  0.747  & 25.6 & 0.297 & 0.236 & 25.4 & 0.289 \\
\bottomrule
\end{tabular}
\pretablecaption
\caption{Comparing with simulated object impact sounds. Lower is better for all metrics.}
\label{Tab:comparing_with_simulation}
\end{table*}

\begin{figure*}
    \centering
    \includegraphics[width=\linewidth]{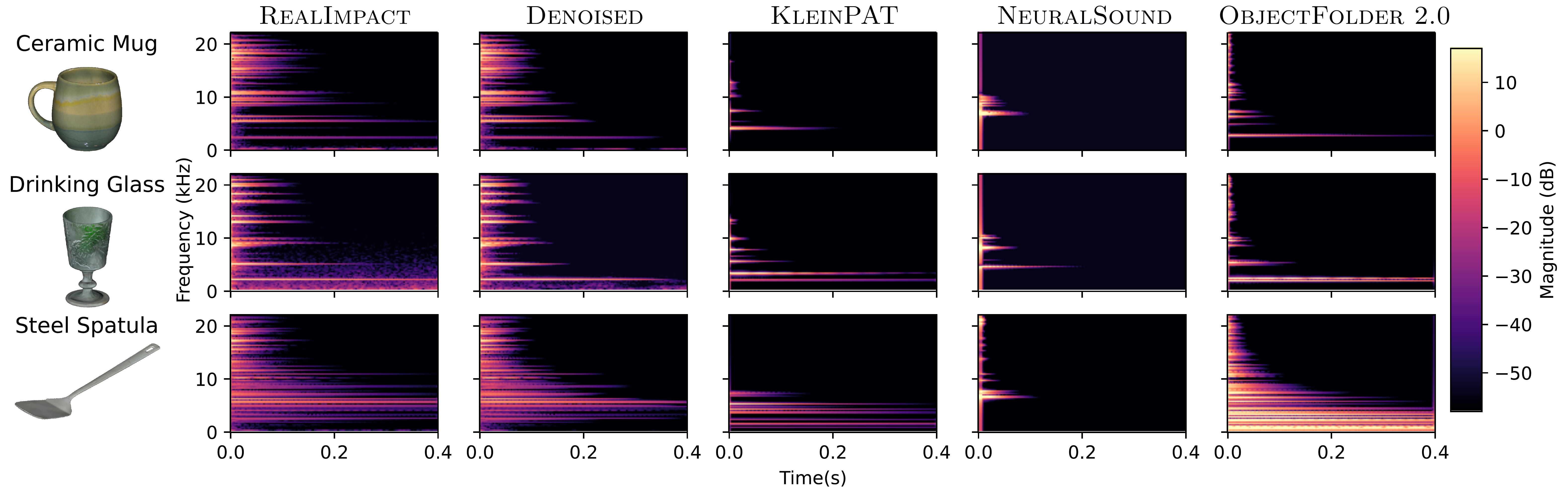}
    \vspace{-3mm}
    \precaption
    \caption{Comparison of spectrograms from our collected recordings versus simulation frameworks. Each spectrogram represents the sound recorded or simulated from a sample point at (8, 22, 13) cm in Cartesian space from the center of the base of the object, and each row corresponds to striking the same vertex on the object.\postcaption}
    \label{fig:sim_vs_real}
\end{figure*}

\subsection{Comparing Simulated and Real Impact Sounds}
\label{sec:baselines}

Our first task is to compare sounds synthesized by existing sound rendering methods to the recordings of~\name in order to demonstrate typical measurement and modeling discrepancies. For this purpose, we ran each baseline method \emph{out-of-the-box} without any attempt to fine-tune its model and/or hyperparameters, including material parameters, such as elastic stiffness (Young's modulus) and damping (e.g., $\alpha$ and $\beta$ in the case of \kleinpat). We also did not unify the finite element analysis representations across different methods, including finite element type (\kleinpat uses first-order tetrahedral elements, whereas ObjectFolder 2.0 uses second-order ones), and tetrahedral meshes. These out-of-the-box comparisons simplify the analysis and highlight the ability to benchmark any existing or new simulation methods given a dataset such as~\name, but exhibit various modeling oversights. We leave the work to narrow the gap between each baseline to the dataset as future work. We provide more conjectures on why these discrepancies exist in the limitations section.

\myparagraph{Baselines.} We provide high-level descriptions of each baseline and refer the readers to Appendix~\ref{app:baselines} or directly to the linked work for more details:
\vspace{-1mm}
\begin{itemize}
    \setlength\itemsep{-1pt}
    \item \textsc{White Noise:} Random noise which has been adjusted to the same loudness as the average loudness of the recordings on a per-object basis.
    \item \textsc{Random Impact Sound:} A random impact sound recording from our dataset.
    \item \kleinpat~\cite{wang2019kleinpat}: The modal analysis is run using first-order tetrahedral elements, and the Far-field Acoustic Transfer (FFAT) maps are done using a one-term ($\nicefrac{1}{r}$) scalar expansion.
    \item \textsc{NeuralSound}~\cite{jin2022neuralsound}: The modal analysis is run using an optimization which is warm-started with the outputs of a 3D sparse U-net on voxelized meshes. The FFAT maps are predicted directly by a ResNet-like encoder-decoder structure. For both steps, we use the pretrained weights.
    \item \textsc{ObjectFolder 2.0}~\cite{gao2022objectfolder}: The modal analysis is predicted by an implicit neural representation trained on simulation data using second-order tetrahedral elements. No acoustic transfer values exist in this baseline so we used $\hat{p}(\boldsymbol{x}; \omega) = 1$ throughout.
\end{itemize}
\vspace{-1mm}

Note that the final three baselines all require material properties of the object as input; we uniformly apply the same parameters from Table 4 of~\cite{wang2019kleinpat} for all objects with the same material label in our dataset. 

\myparagraph{Metrics.} We evaluate using the following metrics:  1) L1 spectral loss, a loss based on taking the average L1 distance between log-magnitude spectrograms of different window sizes (used for impact sounds in~\cite{clarke2021diffimpact}); 2) envelope distance, which measures the distance between two audio samples' envelopes over time (used for spatial audio in~\cite{morgadoNIPS18}); and 3) CDPAM~\cite{manocha2021cdpam}, a learning-based perceptual distance metric trained from human judgments of detectable differences between clips.

To mitigate the effects of measurement noise in our evaluation, we compare each baseline both to our deconvolved recordings and to denoised versions of our recordings, which have been denoised with the algorithm of~\cite{sainburg2020finding}. Whereas many denoising algorithms are optimized for human speech, this algorithm has been optimized and validated against broader categories of audio signals from nature. Comparisons of example spectrograms and their denoised counterparts are shown in Appendix~\ref{app:denoising}.

Quantitative results are shown in Table~\ref{Tab:comparing_with_simulation}, and qualitative examples are shown in Fig.~\ref{fig:sim_vs_real}, comparing our recordings of real impact sounds with the simulated sounds using methods from ~\cite{wang2019kleinpat,jin2022neuralsound,gao2022objectfolder}. The \kleinpat baseline performs best according to a spectral loss, whereas \neuralsound performs best according to the perceptual CDPAM loss. Both of these baselines significantly outperform \objectfolder, suggesting that explicitly modeling the acoustic transfer of objects rather than merely their structural vibrations is essential for achieving realism. A random impact sound only outperforms baselines in Envelope Loss when the recording has not been denoised. Each baseline other than white noise performs better on all metrics when compared against denoised versions of our recordings, suggesting that our raw recordings have non-negligible measurement noise, which must be accounted for in future comparisons.

\subsection{Listener Location Classification}
\label{sec:listener_location_classification}

Identifying the location of the listener with respect to the sound source is of great practical interest to many applications in virtual reality and robotics~\cite{rascon2017localization,chen2021waypoints,rajguru2022sound}. In this task, we want to identify the microphone position (angle, height, or distance) from the impact sound recording. 

For each impact in \name, we have the recordings of the impact sound from 600 different listener locations collected from 10 different angles, 15 different heights, and 4 different distances, as illustrated in Fig.~\ref{fig:recording_illustration}.

Particularly, we set up three separate classification subtasks: 1) angle classification, where the goal is to classify the sound into the 10 angle categories (0$^{\circ}$, 20$^{\circ}$, $\ldots$, 180$^{\circ}$); 2) height classification, where the goal is to classify the sound into the 15 height categories, each corresponding to the height of our 15 microphones;
and 3) distance classification, where the goal is to classify the sound into the 4 distance categories (0.23 m,  0.56 m, 0.90 m, and 1.23 m).

For each subtask, we split 90/10 percent of impact sound recordings of an object into the train/test set, respectively.  We train a ResNet-18~\cite{he2016deep} network that takes the magnitude spectrogram of the impact sound as input to predict the angle, height, or distance category. Table~\ref{Tab:listener_location_classification} shows the results averaged across all 50 objects. We observe that predicting height is comparatively easier. We suspect that differences in height strongly influence the spectral details for easier classification.

\begin{table}[t!]
\centering
\begin{tabular}{lccc}
\toprule
 & Angle & Height & Distance   \\
\midrule
\textsc{Chance}  & 10.0 & 6.7 & 25.0 \\ 
Ours  & \textbf{57.9} & \textbf{60.7} & \textbf{67.4} \\ 
\bottomrule
\end{tabular}
\pretablecaption
\caption{Listener location classification results. We report the accuracy (in \%) for angle, height, and distance classification, respectively.\postcaption}
\label{Tab:listener_location_classification}
\end{table}

\subsection{Visual Acoustic Matching}
\label{sec:visual_acoustic_matching}

The ability to match a source sound with the correct corresponding visual input plays an important role in tasks such as speech and speaker recognition \cite{torfi20173d, lee2021looking} or object and event localization \cite{hu2020discriminative, wu2019dual}.
This task aims to match a sound recording with the correct corresponding image. We set up this matching task as binary classification.

For 20 of our objects, we have a total of 200 RGBD images taken simultaneously with our audio recordings, which are collected at a fixed height from the 10 different angles and 4 different distances from which we took audio recordings for each of the 5 different vertices. We generate positive pairs by pairing each sound recording to an image taken at the corresponding angle and vertex. The height and distance of the image are fixed, so there are 50 possible images that the sound recordings correspond to. The distance of the paired RGBD image is selected such that the image captures the position of both the object and the impact hammer. Negative pairs are generated by pairing sound recordings with images that are not at the correct angle and vertex.

We randomly select two heights to be held out for validation and test sets, while the remaining 13 heights are used for the train set. We train an audio-visual network with a ResNet-18 backbone for both the image and audio streams. The network takes in an RGB image as visual input and an impact sound recording as audio input. A fusion layer combines the audio and visual information, and a final fully-connected layer is used to extract audio-visual features for binary classification. 
Table~\ref{Tab:visual_acoustic_matching} shows the quantitative results of the visual acoustic matching task averaged across 20 objects, and we show example inputs and outputs in Appendix~\ref{app:visual_matching}.

\begin{table}[t!]
\centering
\begin{tabular}{lcc}
\toprule
 & Accuracy $\uparrow$ & RMSE $\downarrow$ \\
\midrule
\textsc{Chance}  & 50.0 & 59.7 \\ 
Ours  & \textbf{75.1} & \textbf{47.7} \\ 
\bottomrule
\end{tabular}
\pretablecaption
\caption{Quantitative results of visual acoustic matching. We report the accuracy results (in \%). RMSE angle error (in degrees) is the root-mean-square error in the difference between the angles of the image and sound recording. $\uparrow$ or $\downarrow$ signify higher or lower values are better, respectively.\postcaption}
\label{Tab:visual_acoustic_matching}
\end{table}

%% file: 07_conclusion.tex
\section{Limitations and Conclusion}

We presented~\name, a first-of-its-kind, large dataset of 150k real impact sounds systematically collected in an acoustically treated room, and demonstrated its several use cases on benchmarking existing simulation algorithms and applications on several auditory and audiovisual tasks.

The microphone array stack used in our measurement process is somewhat coarse to capture high-frequency details (e.g., the 12~cm microphone spacing in elevation roughly corresponds to the wavelength of 3~kHz). The angular resolution is chosen at only 20~degrees and can result in aliasing and distortion in the otherwise symmetric radiation fields, as shown in~\S\ref{sec:spatial_sampling}. The diversity of the recorded objects is restricted by the size and load capacity of our supporting thread mesh, and the microphone stack arm. The material descriptions of the objects are artificially lumped  into the categories defined in previous work~\cite{wang2019kleinpat}, but they may not describe the diversity of real-world materials (e.g., different kinds of steels will have different mechanical properties that might affect stiffness and thus the frequency distributions). Future work should look at more efficient ways of capture and sample a wider range of objects.

The comparison of real impact sounds to those generated by current simulation methods exhibit various discrepancies. Many things can lead to the gap between simulations and real recordings: object scanning resolution and reconstruction accuracy, material stiffness and damping parameters, finite-element analysis differences (e.g., element type), and insufficient meshing resolution. 
Also we did not explicitly model hollow objects in the comparisons despite some of our objects being hollow and/or thin, and
contact damping models are missing, which can affect the perceived damping rates and thus the envelope accuracy. As a result, many of the out-of-the-box simulation models' vibrational frequencies do not agree, let alone their estimates of spatialized sound amplitudes, etc. Many of these oversights will affect the comparison ``fairness" but it also demonstrates a significant benefit of the \name dataset: for the first time, we can measure these models using the same yard stick. We hope that this dataset provides future incentives to improve not just the simulation methods and their usage, but also the capture process and datasets that can move the field forward in a significant way.

%% file: 08_appendix.tex
\appendix
\setcounter{figure}{6}
\setcounter{table}{3}
\section*{Appendix}

\noindent The supplementary materials consist of:
\begin{enumerate}
\setlength{\itemsep}{0pt}
    \item[A.] Details of the recording environment.
    \item[B.] Details of the recording apparatus.
    \item[C.] Additional details and observations on the nature of the hammer impacts in our dataset.
    \item[D.] Additional results on interpolating from transfer maps.
    \item[E.] Additional results from measuring the repeatability of measurements from objects of different materials.
    \item[F.] Additional details on our simulation baselines and their assumptions.
    \item[G.] Additional examples of denoised clips from our dataset.
    \item[H.] Example inputs and outputs from our visual acoustic matching task.
\end{enumerate}

\section{Recording Environment}
\label{app:environment}
In order to validate the recording environment and the efficacy of its acoustic treatments in reducing reverberations, we measured the room impulse response with the following procedure. We positioned a loudspeaker at the same location of the room at which we had positioned our objects during our recordings. We then played a ten-second logarithmic sinusoidal sweep from 20~-~20~kHz through the loudspeaker and recorded it with the microphone array. The gantry moved the microphone array through the same positions at which we had collected the object recordings for the dataset, and we recorded the sweep from each position. In this way, we could capture the specific impulse response at every potential recording position to ensure there was good uniformity of the environment and the recordings across all measurement positions we had used for the dataset. We converted each microphone recording of the sine sweep to an impulse response using deconvolution, and then calculated the octave-band reverberation time for the sound to decay by 60~dB (RT60) using the Schroeder method~\cite{schroeder1965new}.

\begin{figure}[t]
    \center
    \includegraphics[width=1\linewidth]{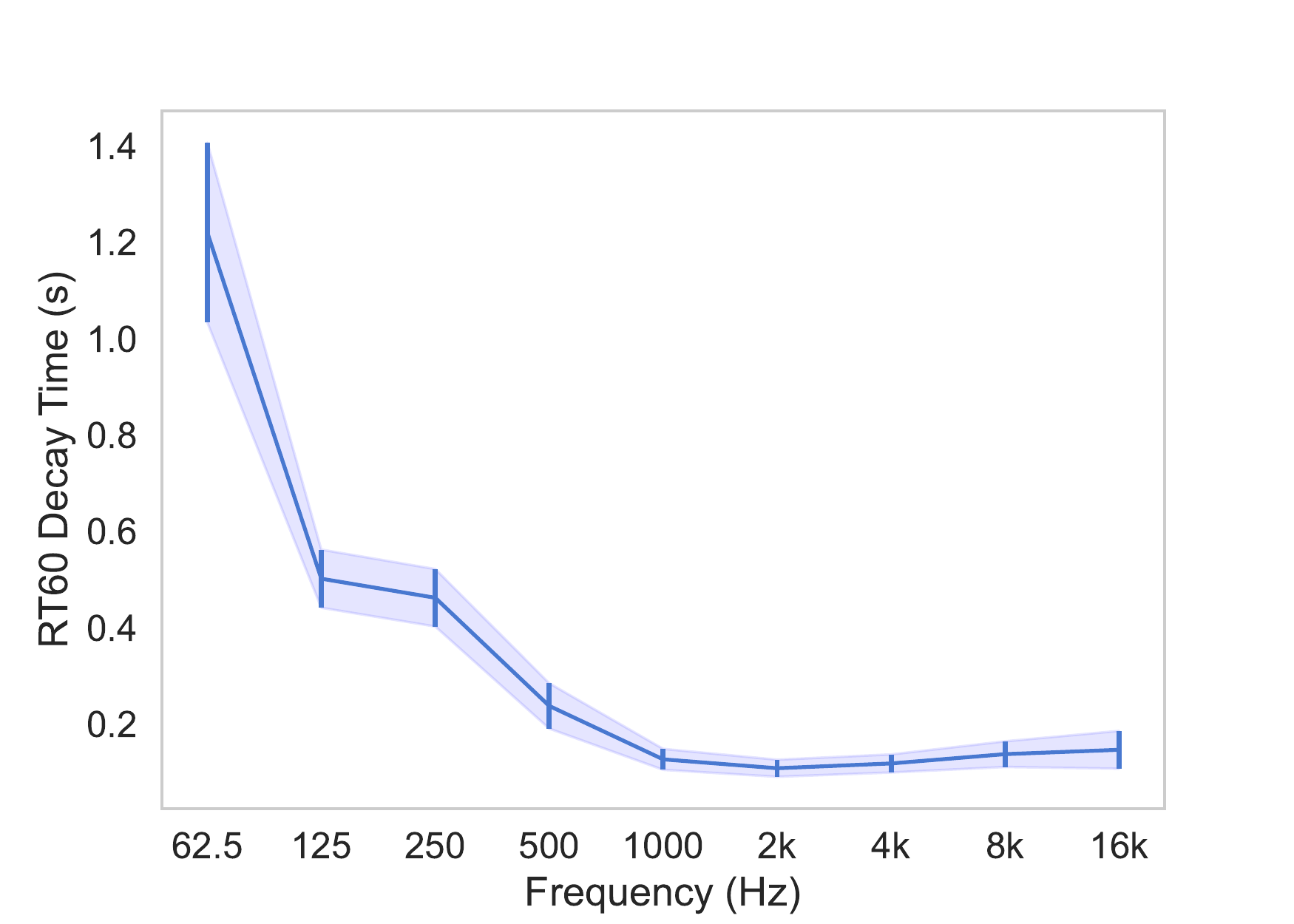}
    \precaption
    \caption{Octave-band RT60 measurements made in the measurement room averaged across all 600 microphone locations.\postcaption}
    \label{fig:RT60}
\end{figure}

The octave band T60 measurements are shown in Figure~\ref{fig:RT60}. The T60 is below 0.2~s for frequencies above 500~Hz, suggesting that the room is fairly anechoic. Below 500~Hz, there is a longer reverb time, as we had made a compromise to treat the room down to a reasonable frequency range while maintaining usable space. With regards to the dataset, since most objects are small, few will have low-frequency resonant modes. Most modes are above 500~Hz, the range in which the room is least reverberant.

\section{Recording Apparatus Details}
\label{app:apparatus}
\begin{figure}[!htb]
    \center
    \includegraphics[width=\linewidth]{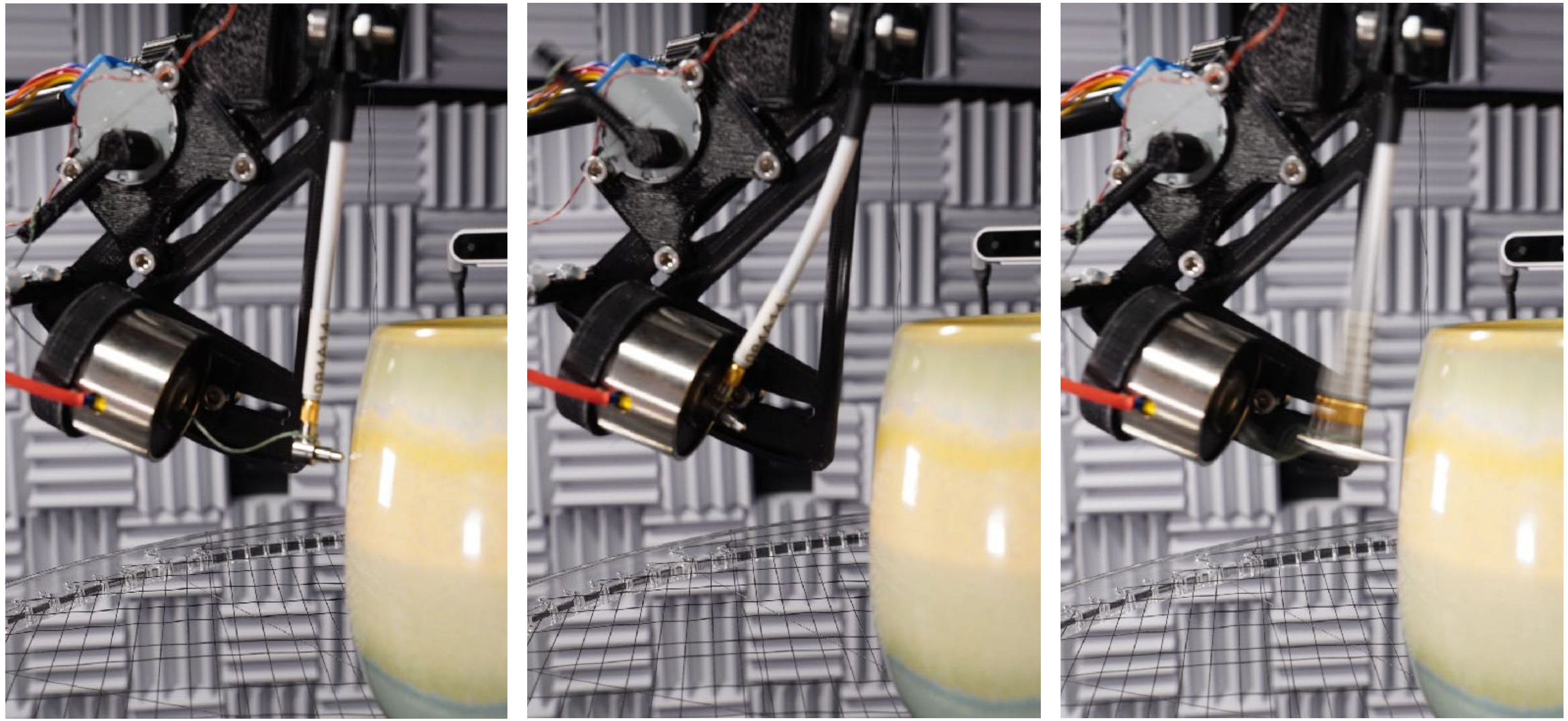}
    \precaption
    \caption{The automated hammer striking mechanism in action. \textbf{(Left)} We manually position the head of the impact hammer such that it is initially near the target impact point without making contact with the object at rest. \textbf{(Center)} To strike the object with the hammer, the motor first winds back the hammer, until it contacts the activated electromagnet to be held into place. The motor then unwinds while the electromagnet holds the hammer. \textbf{(Right)} Finally, the electromagnet releases the hammer to strike the object with as little noise from motion as possible.\postcaption}
    \label{fig:hammer_sequence}
\end{figure}

Here we provide more details and explanations of the mechanical design of our recording apparatus. 

Figure~\ref{fig:hammer_sequence} shows the motion of the hammer striking mechanism. The hammer is cantilevered to the striking apparatus by the end of its handle. The handle consists of a light plastic tube with enough elasticity to store spring energy as the head of the hammer is pulled back to the electromagnet. Furthermore, because this handle is light, the inertia of the system is low enough to mitigate the risk of the head of the hammer bouncing off the object multiple times and polluting our recordings. To further ensure that we do not capture multiple hits in our recordings, we also programmatically validate the recorded signal after each recording.

\begin{figure}[t]
    \center
    \includegraphics[width=0.95\linewidth]{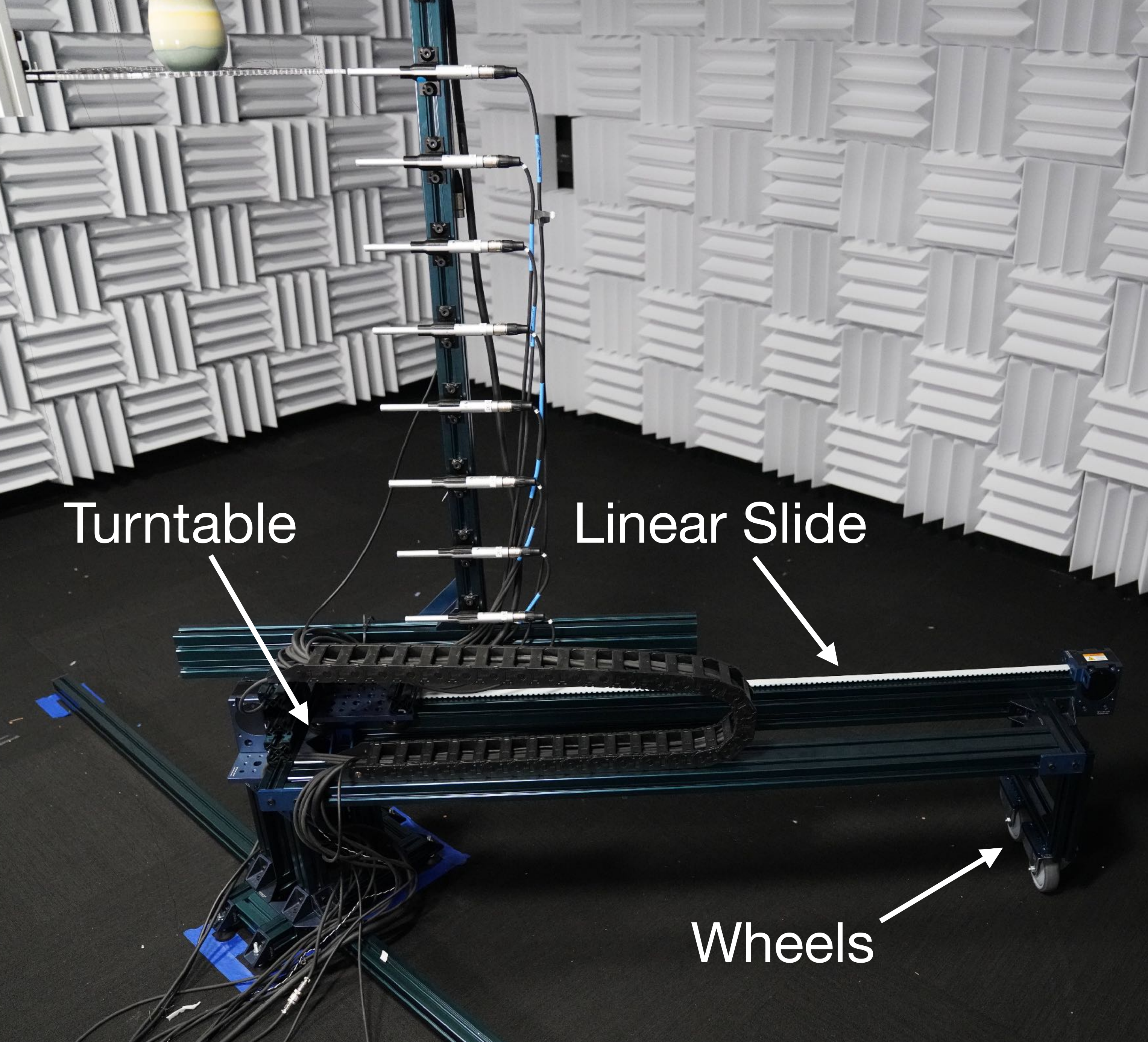}
    \precaption
    \caption{The motion components of the microphone gantry. For rotational motion, a stepper motor rotates a turntable at the center of the gantry, while passive wheels support the other end of the gantry and follow a circular path on the floor. For linear motion, a separate stepper motor drives a linear slide with a timing belt to precisely position the column of microphones.\postcaption}
    \label{fig:gantry_motion}
\end{figure}

Our gantry's motion elements are shown in Figure~\ref{fig:gantry_motion}. The base of the gantry essentially consists of a linear slide resting on a Vention turntable located at the center of rotation, with passive fixed caster wheels at the other end. A stepper motor with a built-in encoder drives the rotation of the turntable. The stepper motor and encoder system have 800 pulses per rotation, and the turntable has a gearbox with a 10:1 gear ratio, meaning that the gantry can theoretically be controlled to 0.045$^\circ$ precision. However, due to the rated backlash of the turntable, the nonzero flexibility of the linear slide and gantry chassis, and the unevenness of the carpet in the room, the gantry may settle into a position up to 1$^\circ$ from where it has been programmed to be for a recording. The linear slide is driven by a separate identical stepper motor and encoder system, with a timing belt moving 150~mm per rotation. With 800 pulses per rotation from the stepper motor, this can theoretically be controlled to 0.19mm precision. The linear slide is more stable at rest and not as susceptible to the unevenness of the carpet in settling to a different position when the motors have been turned off. Because of the flexibility of the column and the mounts of the microphones, we estimate that effective precision of the linear motion of the gantry is 1~mm.

\section{Hammer Impacts}
\label{app:hammer}
\begin{figure*}[t]
    \center
    \includegraphics[width=0.99\linewidth]{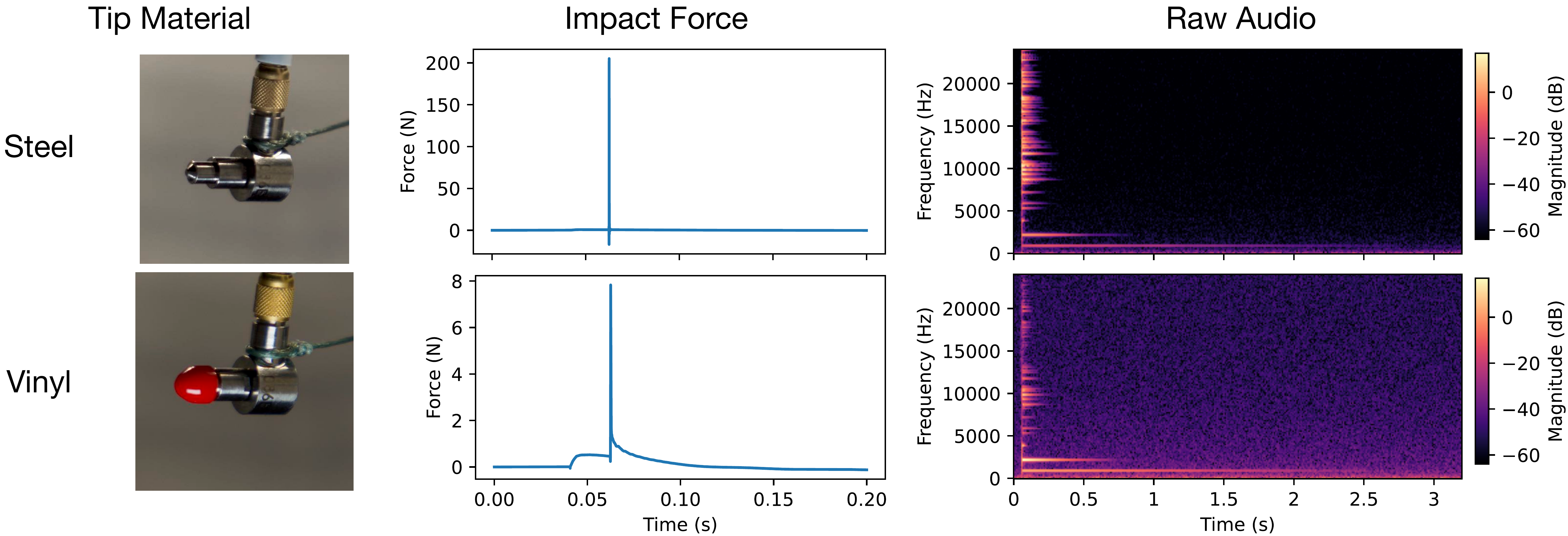}
    \precaption
    \caption{Comparing the resulting forces and audio of striking the ceramic bowl with different materials of impact hammer tip. The top row shows the results of using the standard steel tip as we used in our dataset. The bottom row shows the results of using the tip covered by the soft vinyl cap shown covering the steel tip of the hammer in the image in the bottom row of the left column.\postcaption}
    \label{fig:tip_material}
\end{figure*}

\begin{figure}[htb]
    \center
    \includegraphics[width=0.99\linewidth]{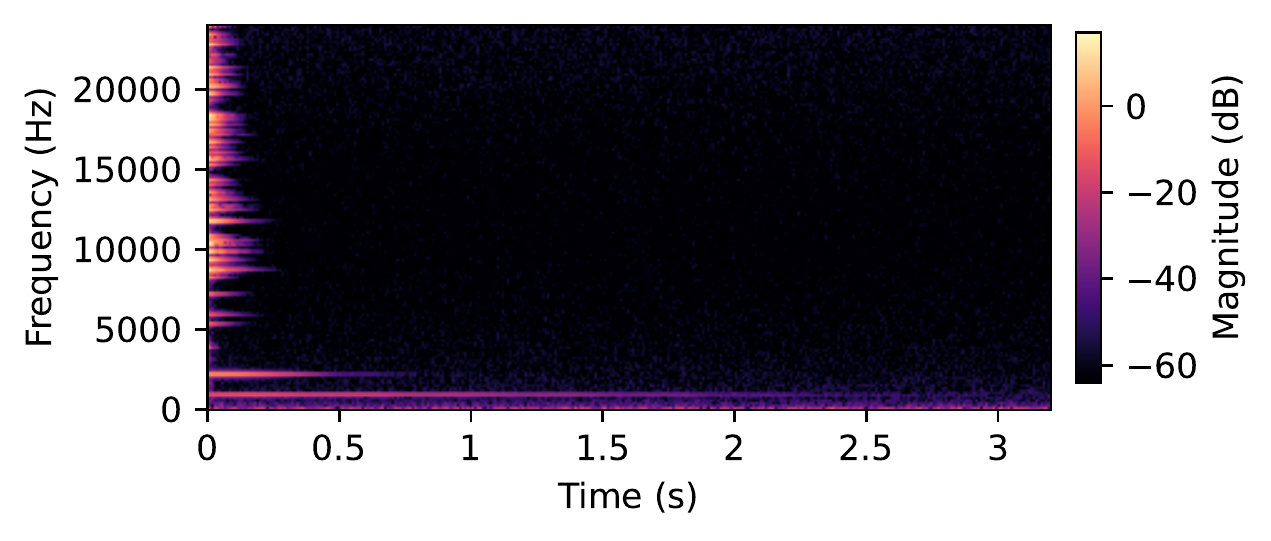}
    \precaption
    \caption{The resulting impulse response estimated by deconvolving the hammer contact forces from the recording of the steel tip striking the ceramic bowl shown in Figure~\ref{fig:tip_material}.\postcaption}
    \label{fig:steel_deconvolved}
\end{figure}

Our using a custom apparatus to strike objects with our steel-tipped impact hammer is important for measuring the contact forces as well as increasing precision and repeatability. We discuss some other implications of these design choices.

\myparagraph{Impact locations}
We choose five striking locations for each object manually, based on multiple trade-offs. First, we generally choose striking locations which optimize for coverage of the different salient regions of each object (\textit{e.g.}, choosing a location on the handle of a mug as well as on the side near its lip). Second, we avoid choosing two striking locations which are symmetric to each other about a plane or axis of symmetry in the geometry of the object. Third, we choose points which are \textit{reachable} by the tip of the hammer, given that the striking apparatus limits the tip's reach. And finally, we choose points which eliminate or minimize the striking apparatus' occlusion of the line of sight between the object and any of the microphones.

\myparagraph{Impact forces}
Though the striking apparatus provides a rather precise and repeatable swinging motion to the hammer, we observe some variation in the striking forces we measured for each object, object vertex, and even vertex trial. The hammer's instantaneous peak striking force is mostly a function of the hardness and restitution of each object's material, ranging from 1.07 to 298 N across the dataset, with a mean of 109 N across all objects. The average standard deviation of the peak forces across all vertices from a single object is 29.8 N, and the average standard deviation across all trials of the same vertex is 11.7 N.

\myparagraph{Hammer material}
The impact hammer is comprised of a plastic handle and a hardened steel tip. The plastic handle emits minimal, but non-negligible, sound as it swings and strikes objects. The tip of the hammer is small enough that its modes of vibration all have frequencies above the Nyquist frequency of our recordings as well as human audible frequencies, thus not directly influencing our impact recordings. The hardened steel tip of the hammer maximimizes the repeatability of impacts and also ensures that impacts are as sharp as possible to both excite the high-frequency modes of each object and make each strike as loud as possible to boost the signal-to-noise ratio of our recordings. This in turn allows us to characterize the impulse response as precisely as possible. Using a softer material for the hammer creates contacts which are longer, which essentially low-pass filters the impulse response of the object~\cite{mali2018study}, and softer, which decreases the signal-to-noise ratio of recordings. In order to demonstrate this, we compare the results of striking the ceramic bowl with the steel tip we used for our dataset, compared to those of striking the bowl with the steel tip covered by a soft vinyl cap in Figure~\ref{fig:tip_material}. Note that for the vinyl-capped tip, the duration of the impact force is indeed longer, and its peak magnitude is much lower. The audio of the impact sound from the vinyl tip is accordingly much quieter, with much more evident noise in the spectrogram confirming a lower signal-to-noise ratio.
\begin{figure}[t]
    \center
    \includegraphics[width=0.95\linewidth]{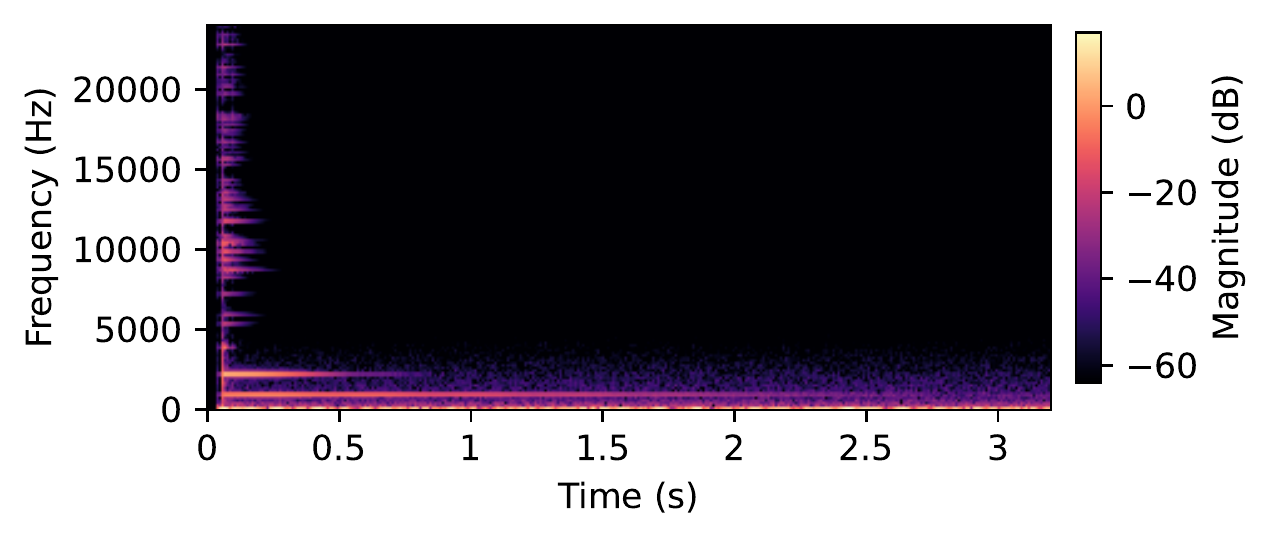}
    \precaption
    \caption{The result of na{\"i}vely estimating the sound of striking the ceramic bowl with the vinyl tip by convolving the impulse response from Figure~\ref{fig:steel_deconvolved} with the impact forces of the vinyl tip shown in the bottom of the middle column in Figure~\ref{fig:tip_material}.\postcaption}
    \label{fig:vinyl_convolved}
\end{figure}

\begin{figure*}[ht]
    \center
    \includegraphics[width=0.96\linewidth]{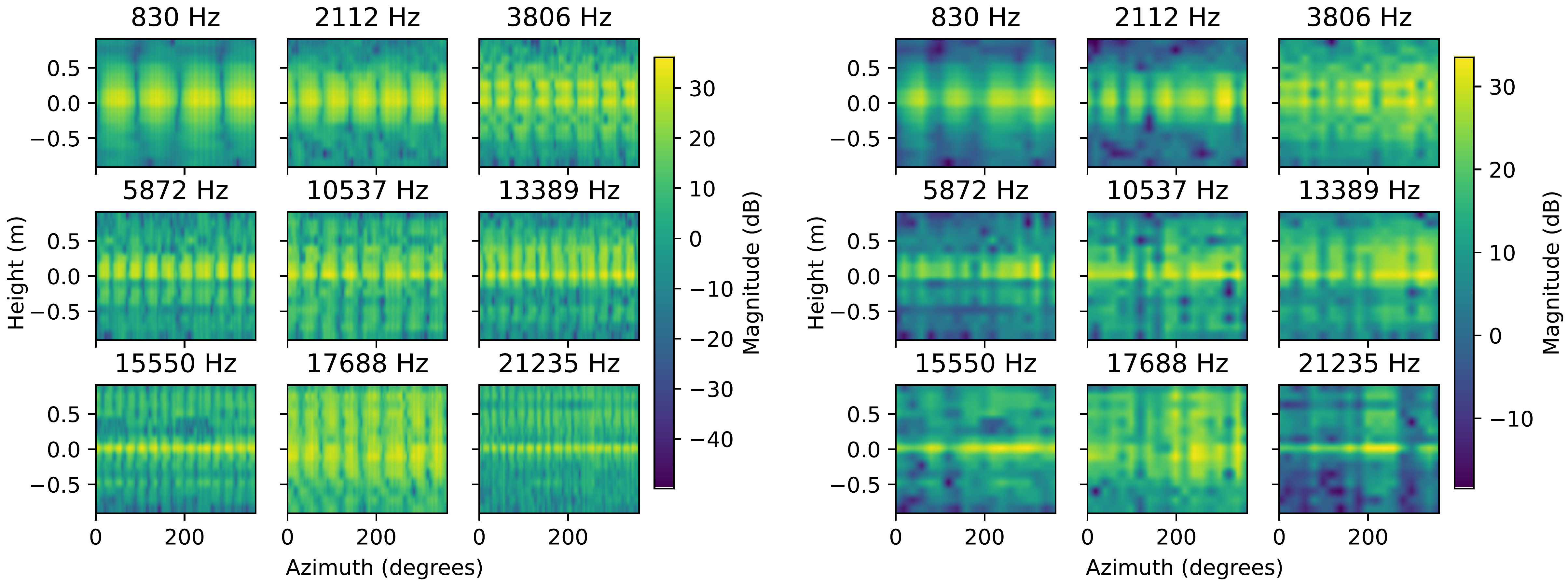}
    \precaption
    \caption{Comparing ground truth measurements versus interpolated mode shape transfer maps of the nine most salient modes of a ceramic bowl. \textbf{(Left)} Ground truth measurements, measured at an azimuth angle resolution of 1$^{\circ}$. \textbf{(Right)} The results of linearly interpolating a 1$^\circ$ azimuth resolution from the 20$^\circ$ resolution measurements used in our data collection process.\postcaption}
    \label{fig:interpolation_map_comparison}
\end{figure*}
\begin{figure}[ht]
    \center
    \includegraphics[width=0.9\linewidth]{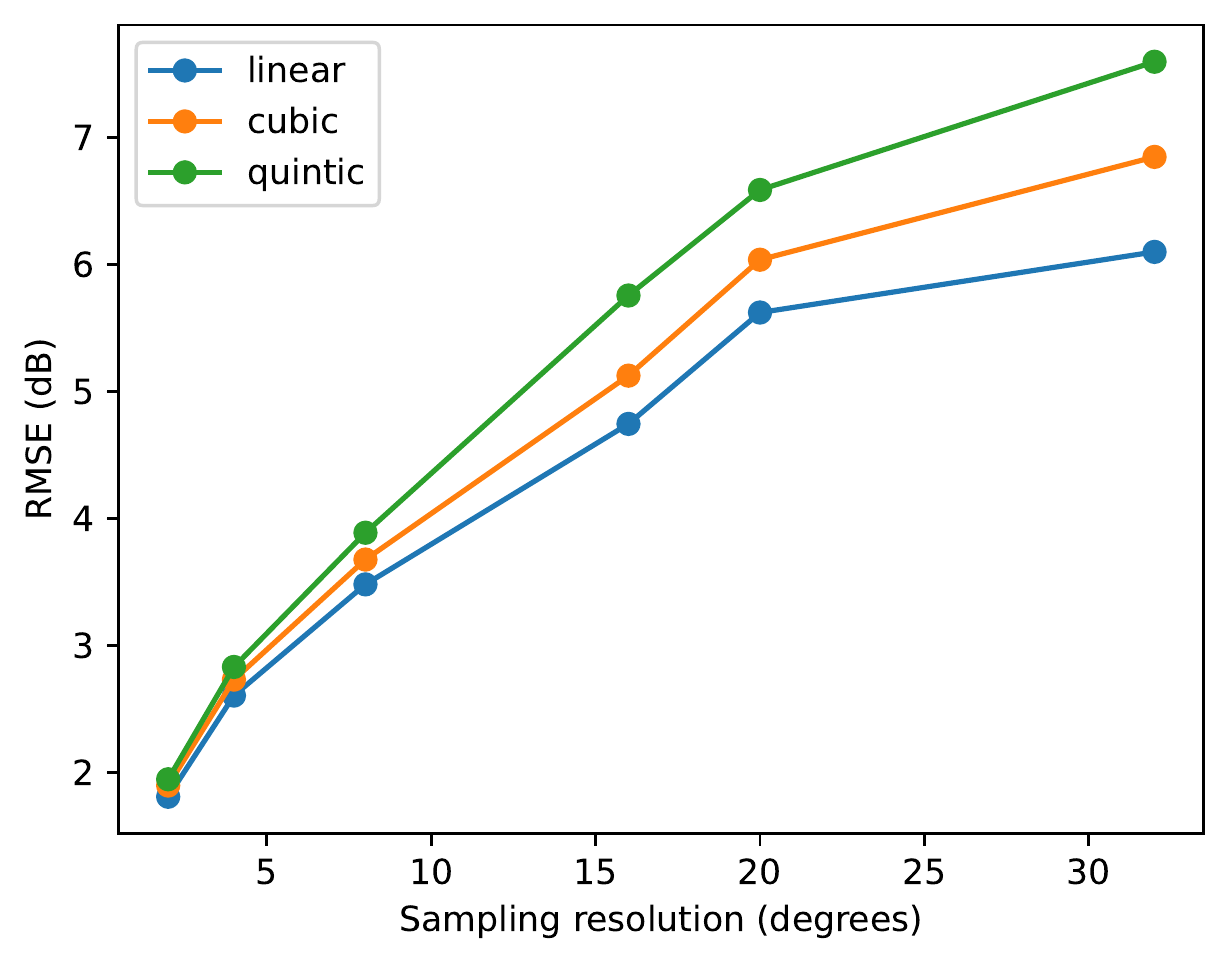}
    \precaption
    \caption{Comparing different interpolation methods for their error in interpolating 1$^\circ$ transfer maps of the ceramic bowl from different levels of azimuth angle coarseness, averaged across the bowl's nine most salient modes.\postcaption}
    \label{fig:interpolation_methods}
\end{figure}

\begin{figure}[ht]
    \center
    \includegraphics[width=0.9\linewidth]{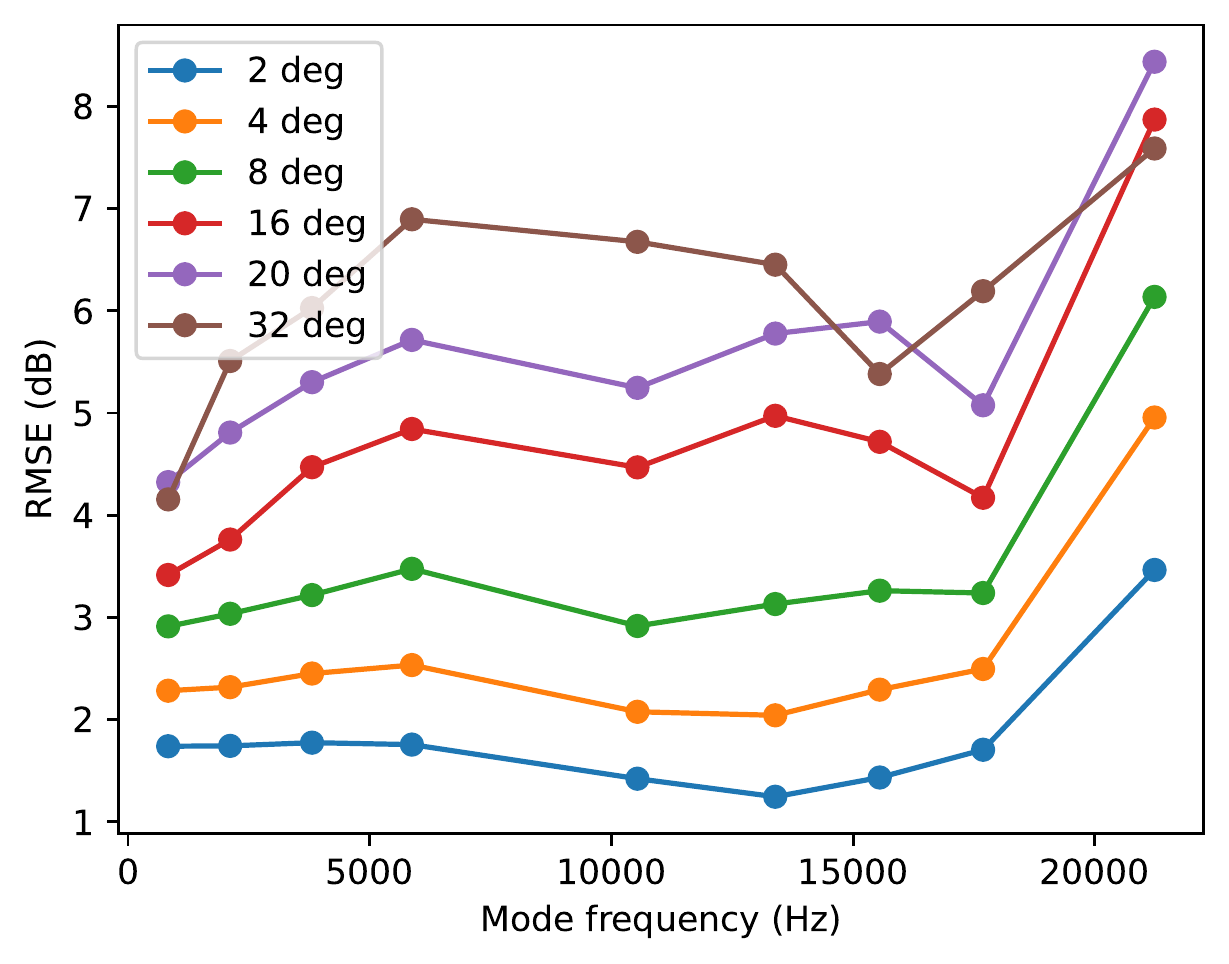}
    \precaption
    \caption{Error of linear interpolation toward estimating transfer maps of 1$^\circ$ azimuth resolution from the ceramic bowl at each mode frequency, stratified by the coarseness of azimuth angle resolution on which each interpolation is based.\postcaption}
\label{fig:linear_interpolation_frequencies}
\end{figure}

However, we can use the deconvolved impulse response from the measurements of impacts using the steel tip to predict the sounds an object would make under different contact conditions, including being struck by a different material. The recording of the steel tip striking the ceramic bowl shown in the top row of Figure~\ref{fig:tip_material} yields the deconvolved impulse response shown in Figure~\ref{fig:steel_deconvolved}. We can then convolve this impulse response with new hammer contact forces to make a na{\"i}ve prediction of the sound the object would make when acted upon by those contact forces. For example, we can use this principle to predict the sound of the ceramic bowl being struck by the soft vinyl tip. We convolve the deconvolved impulse response from the steel tip with the contact forces from the vinyl tip, with the result shown in Figure~\ref{fig:vinyl_convolved}. When compared to the ground truth audio recorded from the impact of the vinyl tip (shown in the spectrogram at the bottom right of Figure~\ref{fig:tip_material}), the prediction shows a modal response with very similar characteristics to that of the ground truth, yet markedly different from the modal response of the steel tip at the top left of Figure~\ref{fig:tip_material}. Further, by using the impulse response from the steel tip with a much higher signal-to-noise ratio, the prediction is less polluted by measurement noise than the actual ground truth recording.

\section{Interpolating from Transfer Maps}
\label{app:interpolation}

Here we show some results from attempting to na\"{i}vely interpolate high-resolution mode shape transfer maps from lower-resolution maps.
First, in addition to those already shown in 
Figure~\ref{fig:transfer_resolution}, the ground-truth high-resolution transfer maps from salient modes of the ceramic bowl are shown on left side of Figure~\ref{fig:interpolation_map_comparison}. These additional transfer maps have also been
collected by the same procedure described in~\S~\ref{sec:spatial_sampling} and processed by the procedure described in~\S~\ref{sec:processing}.

\begin{figure*}[t]
    \center
    \includegraphics[width=0.49\linewidth]{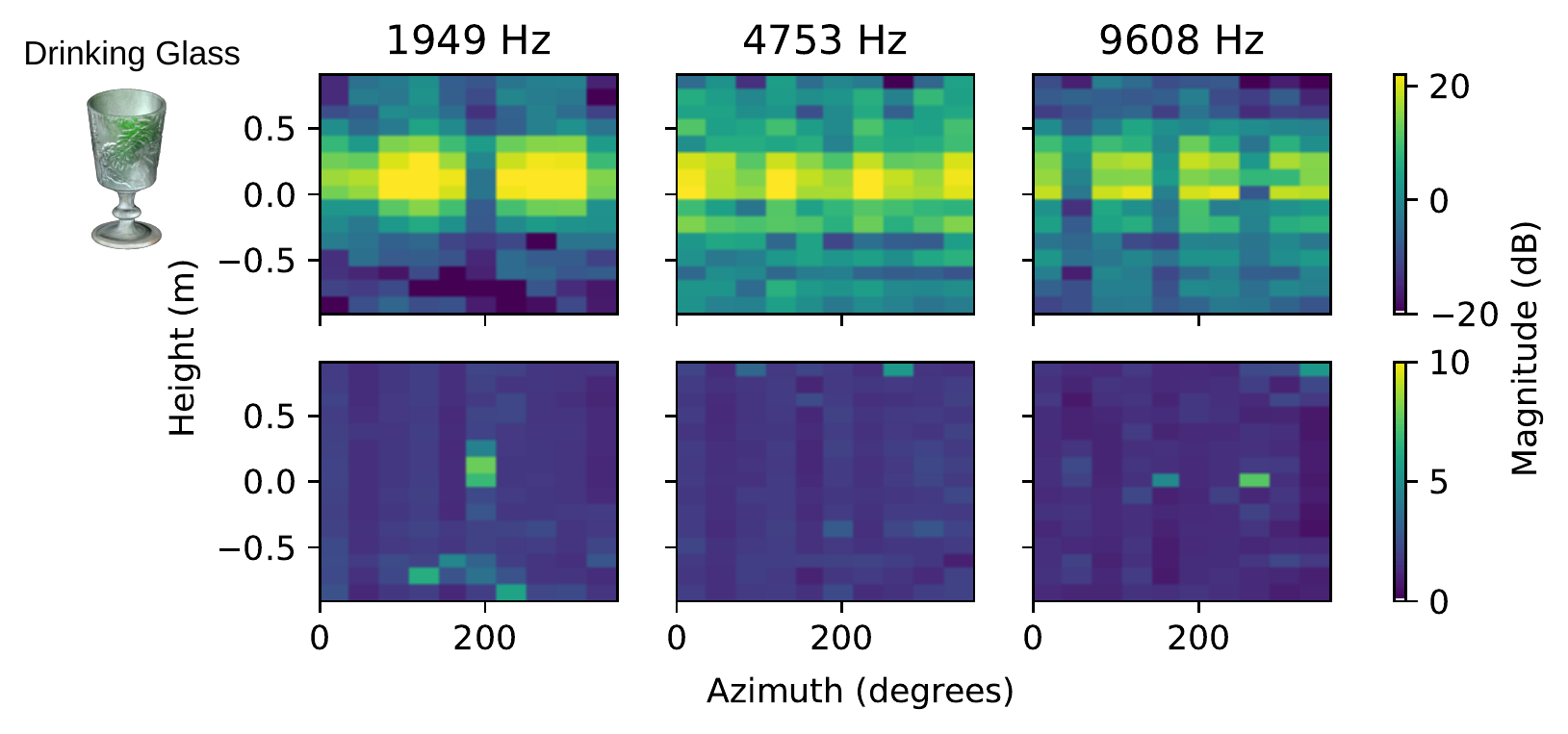}
    \vspace{-2mm}
    \includegraphics[width=0.49\linewidth]{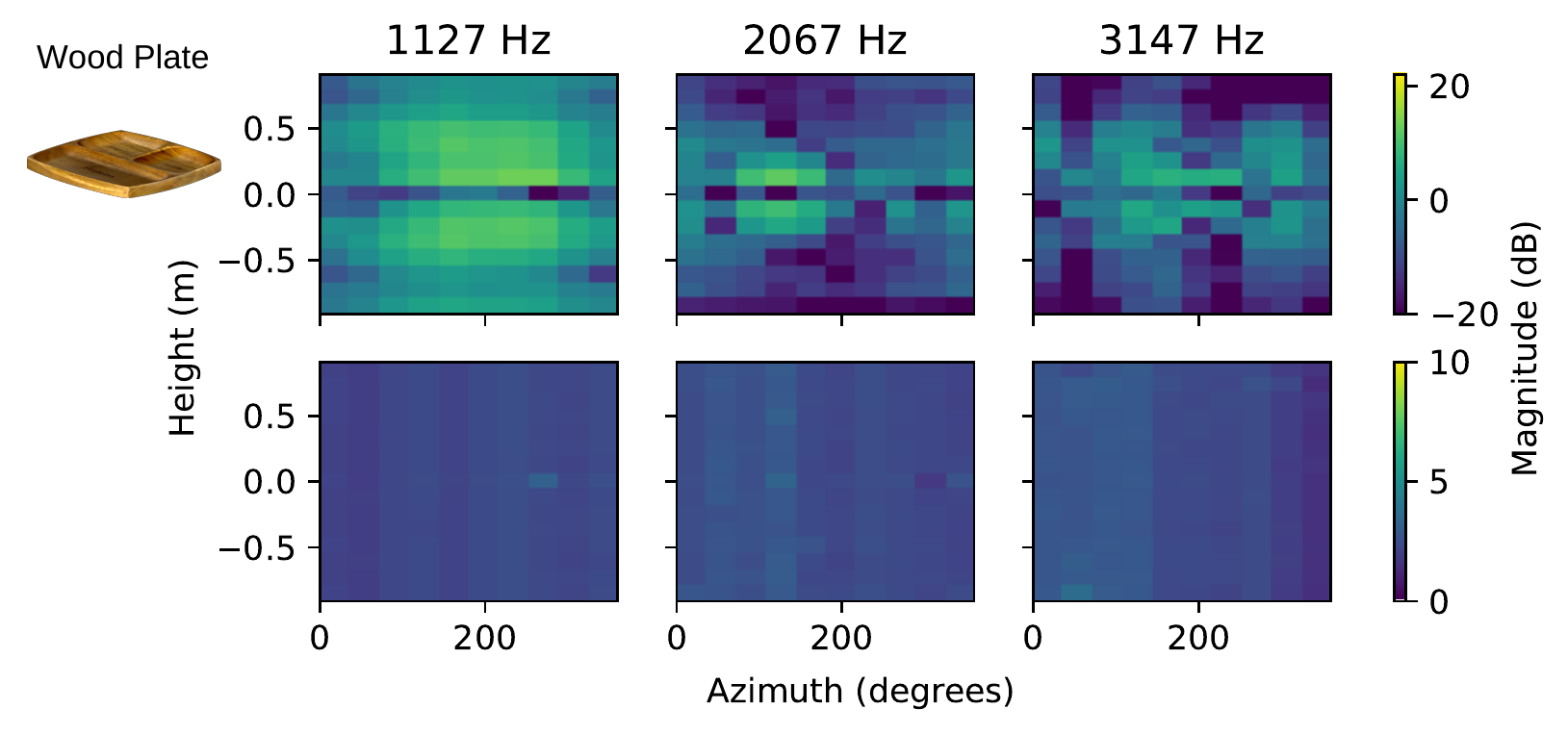}\\
    \includegraphics[width=0.49\linewidth]{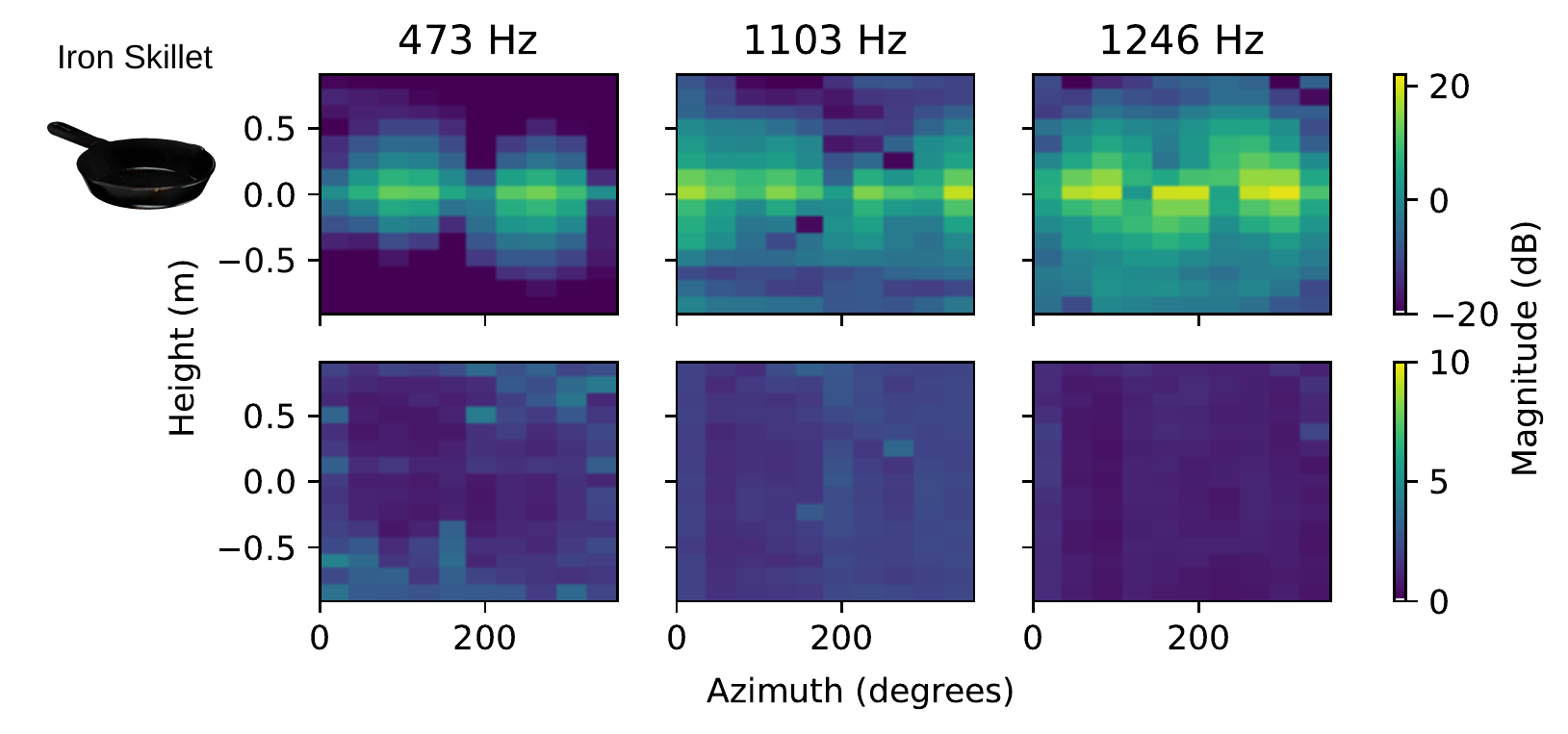}
    \vspace{-2mm}
    \includegraphics[width=0.49\linewidth]{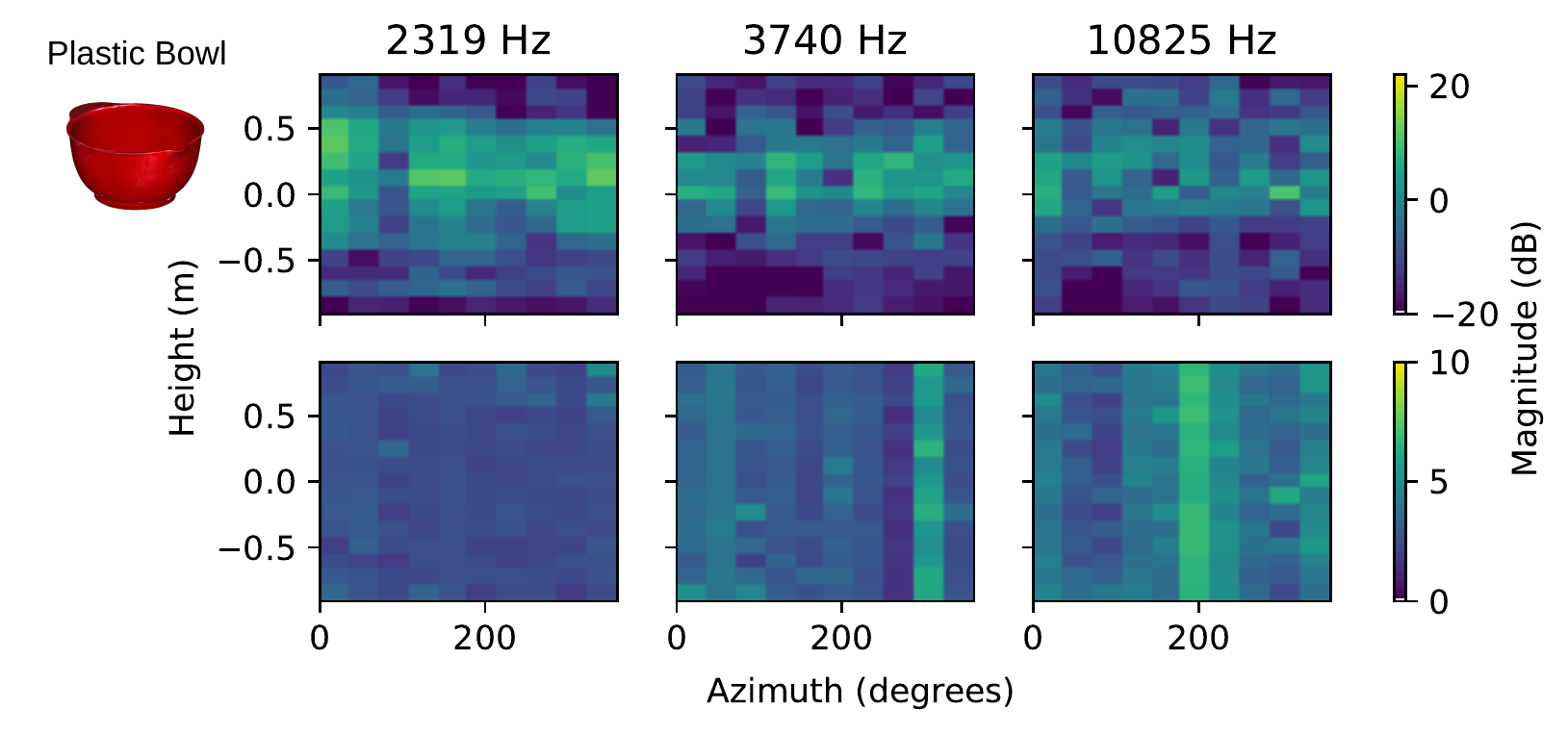}\\
    \includegraphics[width=0.49\linewidth]{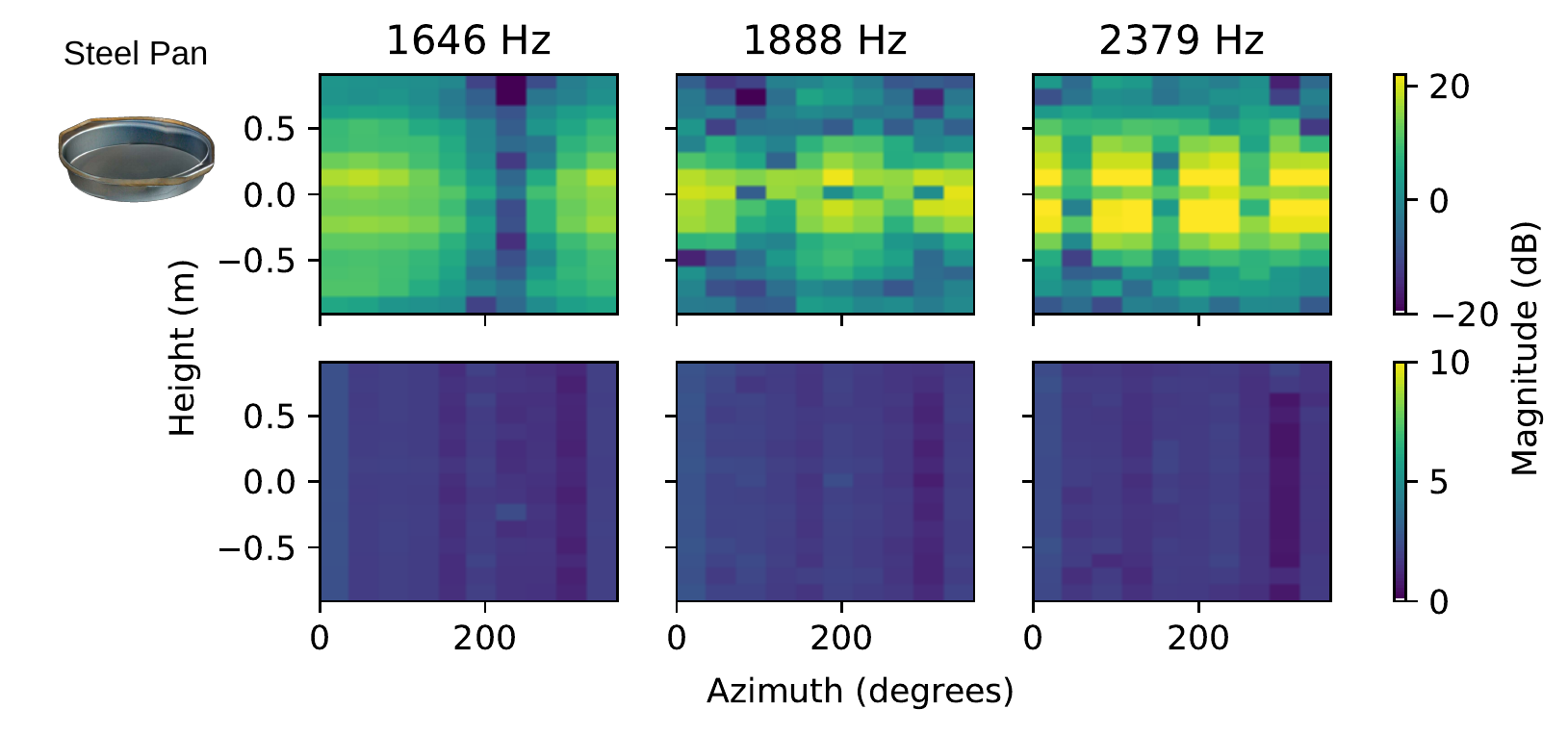}
    \includegraphics[width=0.49\linewidth]{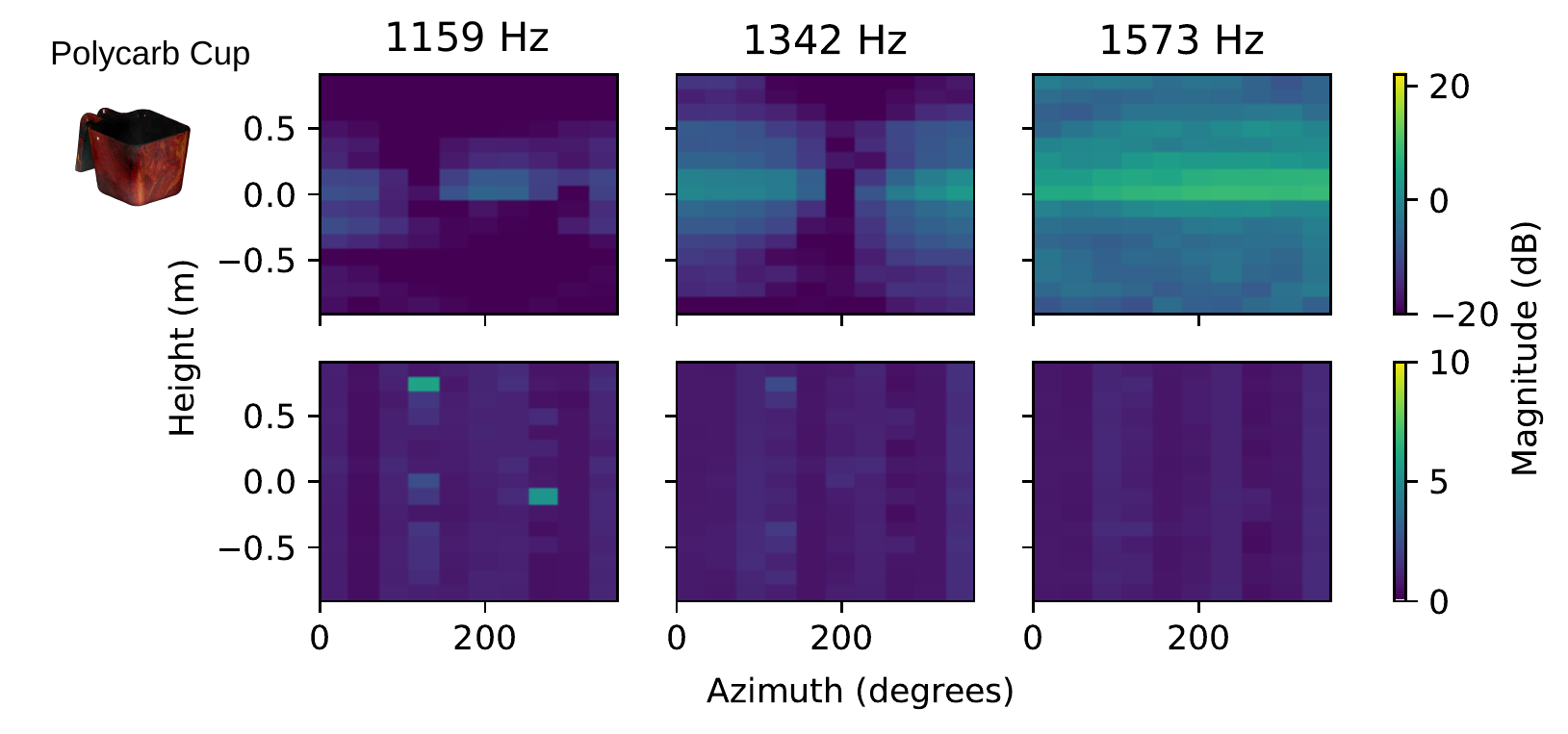}\\
    \precaption
    \caption{Measuring repeatability of our measurements by visualizing transfer maps of vibrational frequencies of the objects of different materials, measured at 23~cm from the center of each object. The top row of transfer maps for each object shows the mean of 10 trials of measurements of striking the same vertex on the object, while the bottom shows the relative standard deviation of the 10 trials.\postcaption}
    \label{fig:transfer_repeatability_additional}
\end{figure*}

We downsample each of these maps to increasingly coarse azimuth angle resolutions and attempt to interpolate back to 1$^{\circ}$ azimuth resolution using linear, cubic, and quintic interpolation methods, then measure the RMSE in decibels of each method at each coarseness of resolution. We average the error of each method across the mode transfer maps of each the nine frequencies and show the results in Figure~\ref{fig:interpolation_methods}. A simple linear interpolation outperforms the cubic and quintic interpolations at every level of coarseness. Figure~\ref{fig:linear_interpolation_frequencies} shows the error of linear interpolation from each coarseness of azimuth angle resolution, with separate error for each mode frequency. The mode shape transfer maps from the 13389, 15550, and 21234 Hz modes suffer the highest errors as the coarseness of sampling increases. As seen on the left of Figure~\ref{fig:interpolation_map_comparison}, the transfer maps for each of these modes have especially high frequency of repetition of their nodes with respect to azimuth angle. Increasing the coarseness of the spatial sampling resolution beyond the Nyquist frequencies of each of these patterns is bound to expand the error of interpolation.

As described in~\S~\ref{sec:data_collection},
our dataset provides sound fields measured at 20$^\circ$ azimuth angle resolution. To demonstrate the challenges of using our dataset to interpolate a sound field, we show the results of linearly interpolating 1$^\circ$ azimuth resolution transfer maps of the ceramic bowl from 20$^\circ$ transfer maps on the right side of Figure~\ref{fig:interpolation_map_comparison}. These results reflect that a na\"{i}ve interpolation, without any domain-specific model bias or priors, will be prone to high errors when trying to interpolate sound fields from the azimuth resolutions at which we have sampled them from our dataset. This motivates future work which is able to use priors to fit high-resolution sound fields from the spatial resolution of the sound fields in our dataset, or perhaps interpolate from an even more minimal amount of measurements.

\section{Additional Repeatability Results}
\label{app:repeatability}
Along with measuring the repeatability of the ceramic bowl (Figure~\ref{fig:transfer_repeatability}), we measured the repeatability of an object from each of the six additional materials according to the same procedure described in~\S~\ref{sec:repeatability}, conducting 10 trials of our measurements striking a single vertex on each object. We show the mean and standard deviations of the transfers we measured at some sample modal vibrational frequencies for each object in Figure~\ref{fig:transfer_repeatability_additional}.

\section{Baseline Details and Assumptions}
\label{app:baselines}
\begin{table*}[t!]
\setlength{\tabcolsep}{5pt}
\begin{tabular}{rccc}
\toprule & \kleinpat~\cite{wang2019kleinpat} & \textsc{NeuralSound}~\cite{jin2022neuralsound} & \textsc{ObjectFolder 2.0}~\cite{gao2022objectfolder}\\
\midrule
\multicolumn{1}{c}{Modal Analysis \& Model} & & & \\
\cmidrule(lr){1-1}
Finite Element Shape &  Tetrahedral & Hexahedral & Tetrahedral \\
\multirow{2}{*}{Finite Element Order} & \multirow{2}{*}{First} & \multirow{2}{*}{First} & \multirow{2}{*}{Second} \\ \\
\multirow{2}{*}{Inference} & \multirow{2}{*}{Precomputed Table} & LOBPCG Optimization & \multirow{2}{*}{Implicit Neural Representation} \\
 & & w/ Neural Warm Start & \\
 \multicolumn{1}{c}{Acoustic Transfer Model} & & & \\
 \cmidrule(lr){1-1}
 Ground Truth Source & Boundary Element Method & Boundary Element Method & N/A \\
 Inference & Precomputed FFAT Map & Neurally Predicted FFAT Map & N/A \\
\bottomrule
\end{tabular}
\pretablecaption
\caption{Comparing assumptions and methods of each baseline model.\postcaption}
\label{tab:baseline_assumptions}
\end{table*}

As stated in~\S~\ref{sec:baselines}, each baseline we evaluated used different assumptions and techniques for simulating sounds. Additional details of the differences in assumptions and methods are detailed below and summarized in Table~\ref{tab:baseline_assumptions}.

\myparagraph{Baseline Modal Models} Each baseline estimates the structural vibrations of objects through finite element-based modal analysis. \neuralsound computes modal analysis by voxelizing objects into hexahedral meshes, whereas \kleinpat and \objectfolder tetrahedralize objects to capture fine geometric features. Both \neuralsound and \kleinpat use first order mesh elements, while \objectfolder uses second order tetrahedra to model the curvature of finite elements during modal vibrations At inference time, \kleinpat estimates the vibrations of the object by directly computing a modal response from the frequencies and gains (\textit{i.e.}, displacements for each mode shape) at each vertex of the mesh from the results of the LU decomposition. \neuralsound trains a sparse U-Net to output vectors which are used as input to the Rayleigh-Ritz method to approximate eigenvalues and eigenvectors. At inference time, the approximated eigenvalues and eigenvectors are quickly optimized using a Locally Optimal Block Preconditioned Conjugate Gradient (LOBPCG) optimization to arrive at the final eigenvalue and eigenvector estimates. \objectfolder uses the eigenvectors estimated by Abaqus~\cite{abaqus2021dassault} to train an implicit neural representation to estimate the modal gains at any contact point on the object. At inference time, the modal response is constructed by using the frequencies estimated by Abaqus and gains predicted by the implicit representation. All baselines use the Rayleigh damping method for estimating the dampings of each mode, based on the same parameters for each material.

\myparagraph{Baseline Acoustic Transfer Models}
While \objectfolder does not model acoustic transfer, \kleinpat and \neuralsound each use different methods for estimating the acoustic transfer of each object. \kleinpat precomputes Far-Field Acoustic Transfer (FFAT) maps from performing \textit{mode conflation} and computing transfer using a finite-difference time-domain (FDTD) wavesolver. \neuralsound computes FFAT maps using a Boundary Element Method (BEM) solver and uses these maps to train a ResNet-like encoder-decoder network to predict FFAT maps for each mode, using the objects' mesh and the mode frequency as input. At inference time, \kleinpat merely uses its precomputed FFAT maps of each mode of an object, while \neuralsound uses its network to predict the FFAT maps to estimate acoustic transfer of each mode.

\section{Additional Denoising Examples}
\label{app:denoising}
\begin{figure}[!htb]
    \center
    \includegraphics[width=1\linewidth]{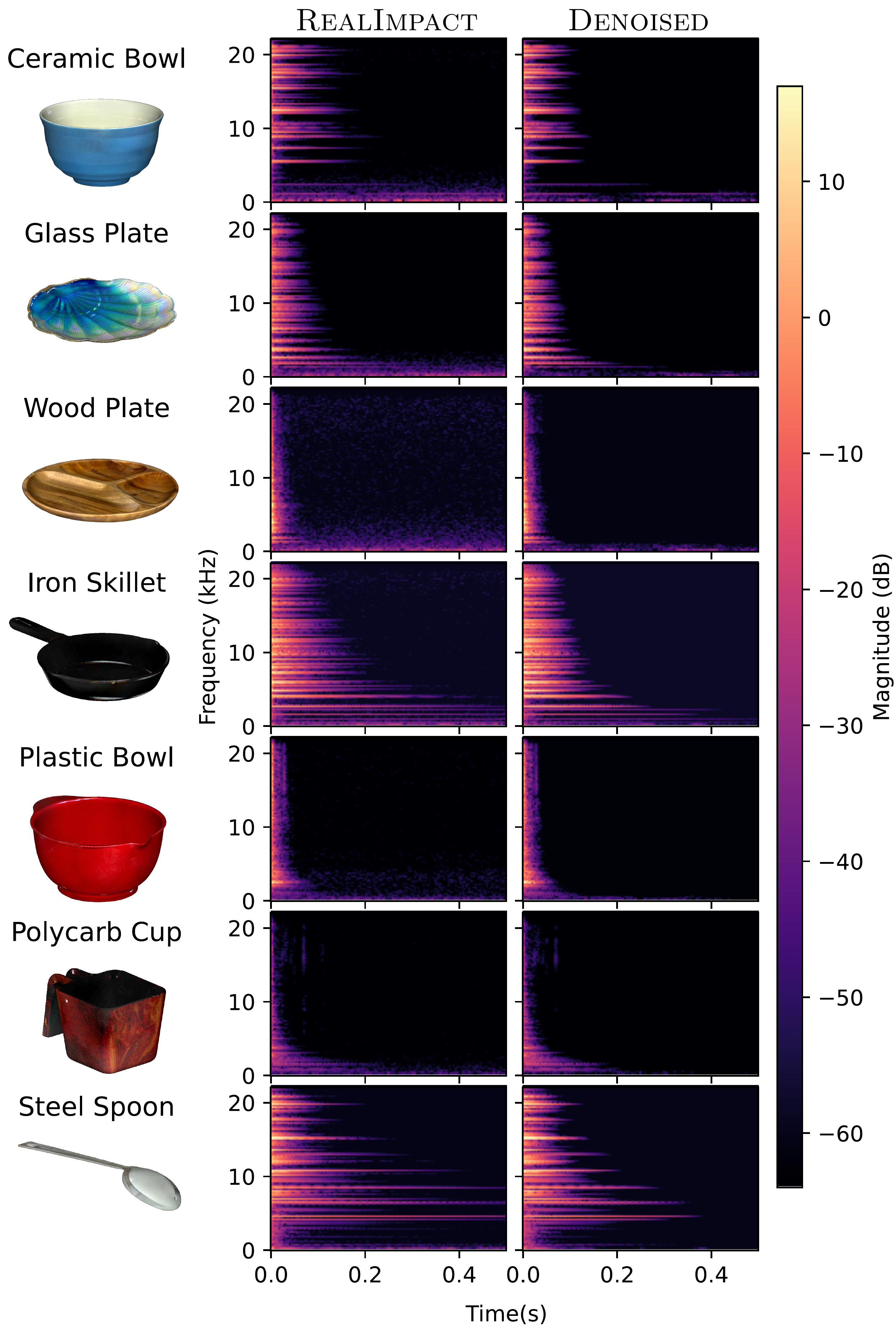}
    \precaption
    \caption{Example spectrograms from \name's raw deconvolved recordings compared to their denoised counterparts, for objects of different materials.\postcaption}
\label{fig:denoising}
\end{figure}
Additional example spectrograms of \name's recordings compared to their denoised versions, produced by the algorithm of~\cite{sainburg2020finding} are shown in Figure~\ref{fig:denoising}. The denoising algorithm seems to be especially helpful in removing the low frequency noise for each object. This is especially evident in the recordings for the ceramic bowl, the glass plate, and the wood plate.

However, while filtering out noise, the algorithm also seems to filter out some important signal. The algorithm filters out modes after they have partially decayed, increasing their effective decay rate. Note in Figure~\ref{fig:denoising} that the modal vibrations of the iron skillet and especially the steel spoon are shortened significantly in their duration by this algorithm. By effectively accelerating the decay of these modes, characterizing the objects' vibrations from these denoised versions could lead to overestimates of the damping properties of the objects and their materials. This motivates future work for an efficient denoising algorithm which is specialized for impact sounds, perhaps inspired by the physics-based principles of modal vibrations, similar to the denoising technique presented in~\cite{clarke2021diffimpact}.

\section{Visual Matching Examples}
\label{app:visual_matching}
\begin{figure}[!htb]
    \center
    \includegraphics[width=1\linewidth]{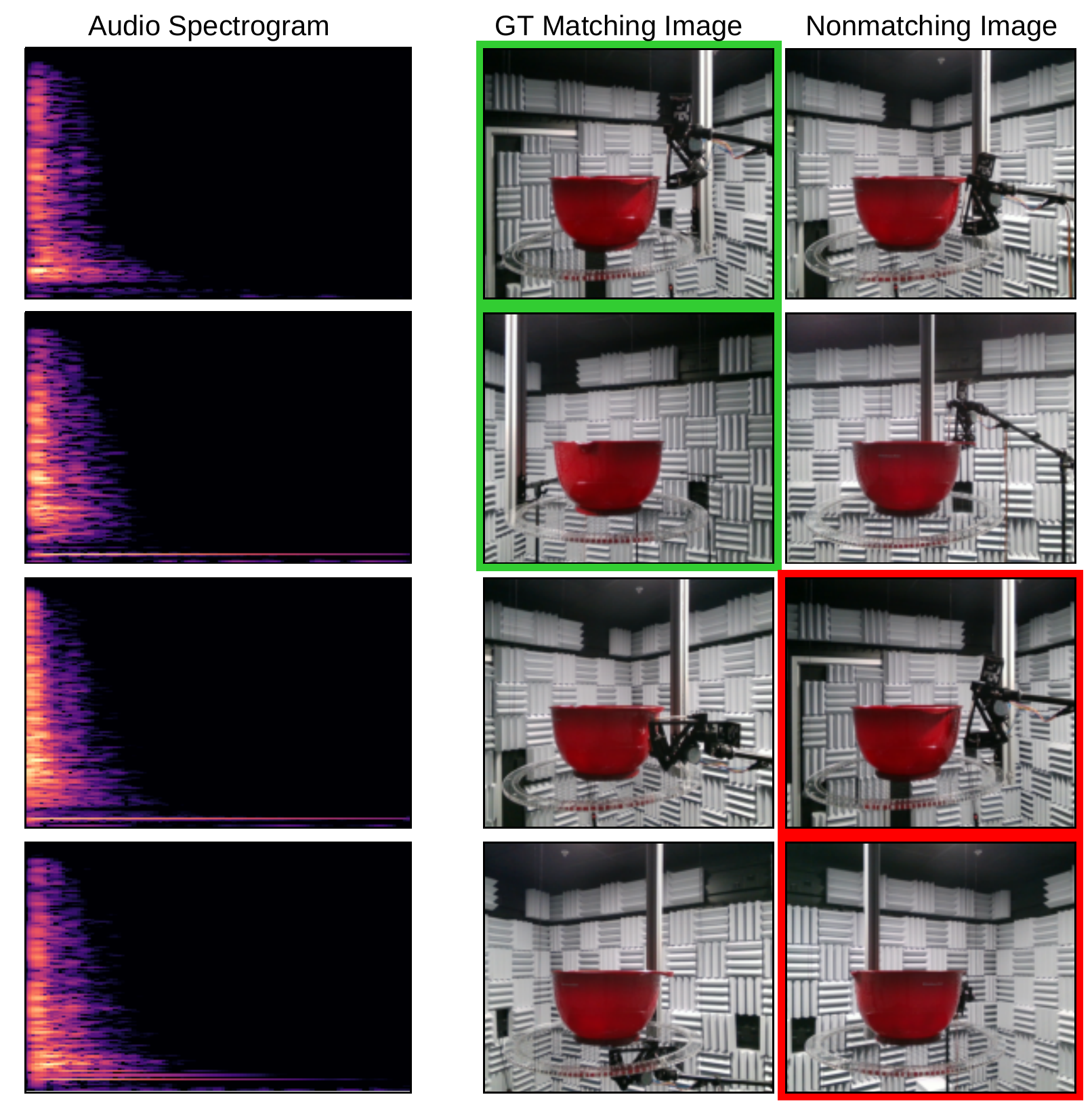}
    \precaption
    \caption{Example success (top two rows) and failure (bottom two rows) cases of our model for the visual acoustic matching task on a \textit{plastic mixing bowl}.\postcaption}
\label{fig:matching_plasticbowl}
\end{figure}
\begin{figure}[!htb]
    \center
    \includegraphics[width=1\linewidth]{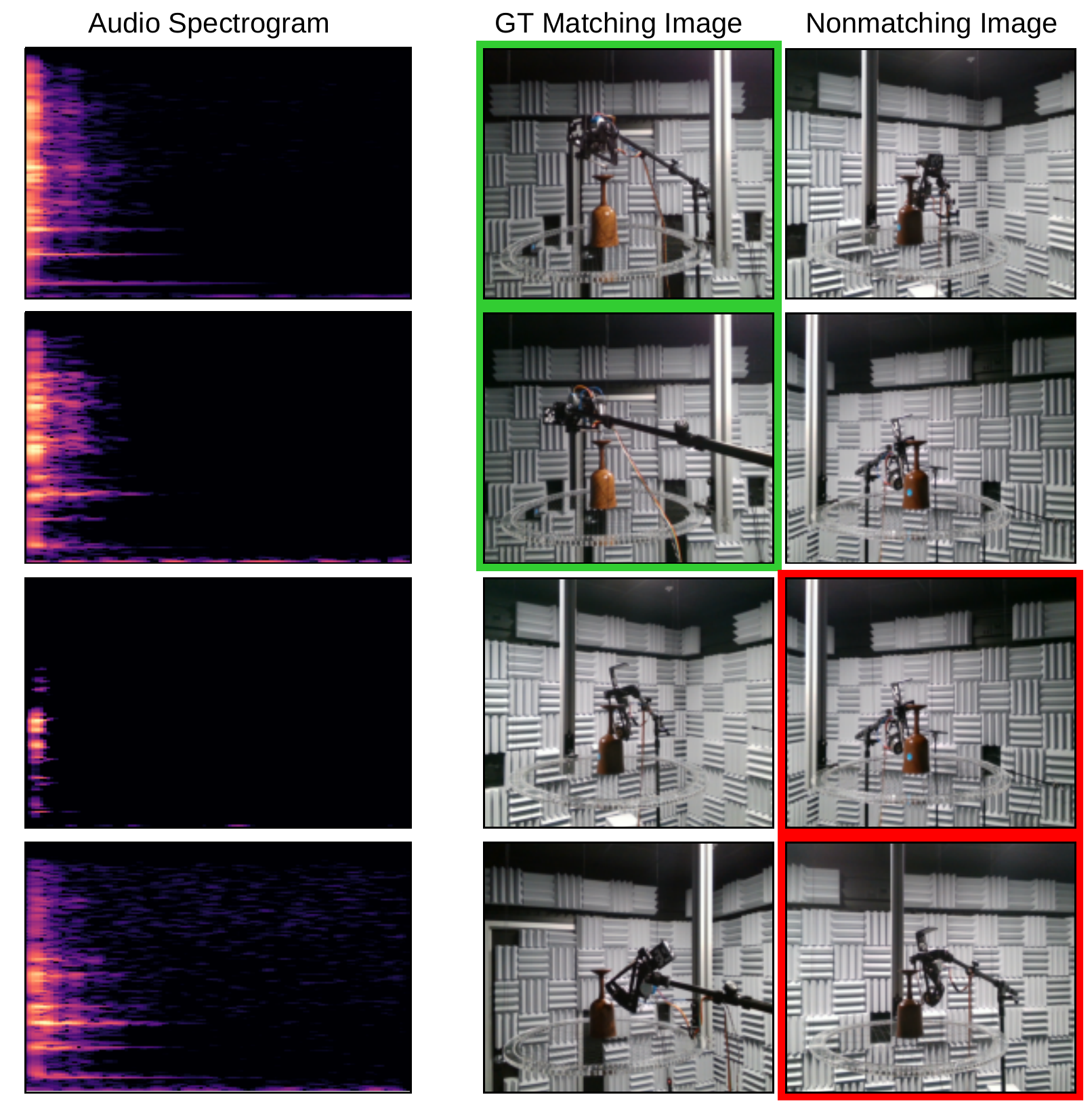}
    \precaption
    \caption{Example success (top two rows) and failure (bottom two rows) cases of our model for the visual acoustic matching task on a \textit{wood wine glass}.\postcaption}
\label{fig:matching_woodwineglass}
\end{figure}
\begin{figure}[!htb]
    \center
    \includegraphics[width=1\linewidth]{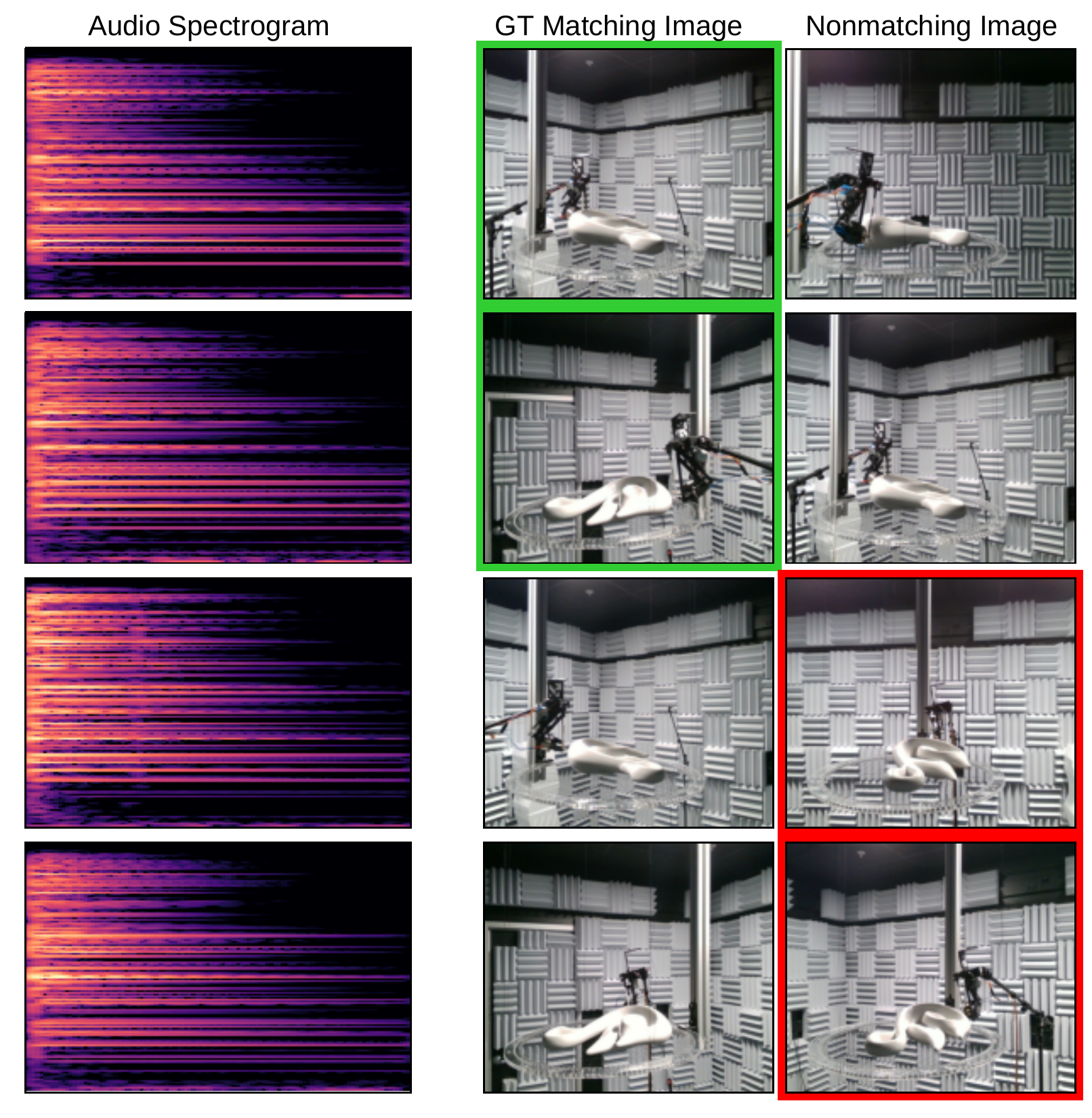}
    \precaption
    \caption{Example success (top two rows) and failure (bottom two rows) cases of our model for the visual matching task on a \textit{decorative ceramic swan}.\postcaption}
\label{fig:matching_ceramicswan}
\end{figure}

Figures~\ref{fig:matching_plasticbowl}~-~\ref{fig:matching_ceramicswan} show a random selection of examples of two success and two failures for three different objects in the visual acoustic matching task described in~\S~\ref{sec:visual_acoustic_matching} of the main manuscript.

For the results of the wooden wine glass shown in Figure~\ref{fig:matching_woodwineglass}, in the success cases, the different position and angle of the hammer stand and object lead to greater visual contrast between the correct matching and nonmatching images in each pair. In both failure cases, the images in each pair appear to be more visually similar to each other. The hammer stand and object are located and angled in similar positions. This contrast in visual similarities and differences between success and failure cases is also evident in the results from the other objects. One important confounding factor is that the model could be exploiting and learning from the visual differences in the room. Each image also captures background details of the recording apparatus and recording room, such as the microphone stand position and patterns in the acoustic padding. It is unclear if the model is learning from the positions of the object and hammer or from other environmental factors in the room. In real-world settings, such external factors may be especially wise to exploit, since they are also likely to influence the acoustic environment of the object and therefore its sound field.